\documentclass[prb,reprint,notitlepage,superscriptaddress,twocolumn,floatfix]{revtex4-1}
\usepackage{amsmath}
\usepackage{graphicx}
\usepackage{subfigure}
\usepackage{lmodern}
\usepackage{amsmath}
\usepackage{braket}
\usepackage{natbib}
\usepackage{tikz}

\usepackage{color}
\usepackage{diagbox}
\usepackage{bm}
\usepackage{amssymb}
\usepackage{tabularx}
\usepackage{hyperref}
\usepackage{tabularx}
\hypersetup{
   bookmarks=true,         % show bookmarks bar?
   unicode=false,          % non-Latin characters in AcrobatÕs bookmarks
 pdftoolbar=true,        % show AcrobatÕs toolbar?
    pdfmenubar=true,        % show AcrobatÕs menu?
    pdffitwindow=false,     % window fit to page when opened
    pdfstartview={FitH},    % fits the width of the page to the window
    pdftitle={My title},    % title
    pdfauthor={Author},     % author
    pdfsubject={Subject},   % subject of the document
    pdfcreator={Creator},   % creator of the document
    pdfproducer={Producer}, % producer of the document
    pdfkeywords={keyword1} {key2} {key3}, % list of keywords
    pdfnewwindow=true,      % links in new window
    colorlinks=true,       % false: boxed links; true: colored links
    linkcolor=magenta,          % color of internal links
    citecolor=cyan,        % color of links to bibliography
    filecolor=magenta,      % color of file links
    urlcolor=magenta           % color of external links
}

\newcommand{\beq} {\begin{equation}}
\newcommand{\eeq} {\end{equation}}
\newcommand{\bea} {\begin{eqnarray}}
\newcommand{\eea} {\end{eqnarray}}
\newcommand{\be} {\begin{equation}}
\newcommand{\ee} {\end{equation}}
\renewcommand{\(}{\left(}
\renewcommand{\)}{\right)}
\renewcommand{\[}{\left[}
\renewcommand{\]}{\right]}

\DeclareMathOperator{\sgn}{sgn}

\newcommand{\ms}{\mathsf}
\newcommand{\bs}{\boldsymbol}

\begin{document}

\title{Chiral Dirac Superconductors: Second-order and Boundary-obstructed Topology}
\author{Apoorv Tiwari}
\affiliation{Department of Physics, University of Zurich, Winterthurerstrasse 190, 8057 Zurich, Switzerland}\affiliation{Condensed Matter Theory Group, Paul Scherrer Institute, CH-5232 Villigen PSI, Switzerland}
\author{Ammar Jahin}
\author{Yuxuan Wang}
\affiliation{Department of Physics, University of Florida, Gainesville, FL 32601}
\begin{abstract}

We analyze the topological properties of a chiral ${p}+i{p}$ superconductor for a two-dimensional metal/semimetal with four Dirac points. Such a system has been proposed to realize second-order topological superconductivity and host corner Majorana modes. We show that with an additional $\ms{C}_4$ rotational symmetry, the system is in an intrinsic higher-order topological superconductor phase, and with a lower and more natural $\ms{C}_2$ symmetry, is in a boundary-obstructed topological superconductor phase. The boundary topological obstruction is protected by a bulk Wannier gap. However, we show that the well-known nested-Wilson loop is in general unquantized despite the particle-hole symmetry, and thus fails as a topological invariant. Instead, we show that the higher-order topology and boundary-obstructed topology can be characterized using an alternative defect classification approach, in which the corners of a finite sample is treated as a defect of a space-filling Hamiltonian. We establish ``Dirac+$({p}+i{p})$" as a sufficient condition for second-order topological superconductivity.
\end{abstract}

\date{\today}
\maketitle

\section{Introduction}
The study of topological phases of matter concerns itself with classifying groundstates of gapped quantum systems and characterizing them via certain robust properties which remain insensitive to adiabatic deformations\cite{Ching-Kai_2016, Ryu_2010, Kitaev_periodic_table, Chen_2013}. A crucial manifestation of topological phenomena is the bulk-boundary correspondence which predicts the existence of non-trivial (ingappable, degenerate or long-range entangled) degrees of freedom on the boundary of a topological phase purely by analyzing the bulk properties.  Paradigmatic examples of such phenomena are the appearance of an odd number of Dirac cones on the surface of the 3D topological insulator\cite{Fu_Kane_Mele_2007}, ingappable helical modes on the edge of the quantum spin hall insulator \cite{Kane_Mele_2005, Kane_2005} and chiral Majorana modes on the edge of the ${p}+i{p}$ topological superconductor\cite{Kitaev_2001} to name a few.    

\medskip \noindent In the past years there have been a gamot of developments that have generalized the bulk-boundary correspondence to include more subtle phenomena. Two classes of generalizations known as {\it{Higher-order topological phases}} and {\it{Boundary-obstructed topological phases}} are of relevance to the present work. Broadly speaking, phases that have a gapped bulk as well as gapped co-dimension-1 boundaries but necessarily support non-trivial degrees of freedom on higher co-dimension boundaries are known as higher-order topological insulators and superconductors\cite{schindler2018higher, schindler2018higher, benalcazar2017quantized, benalcazar2017electric, you2018higher, kunst2018lattice, song2017d, geier-2018,wang2018weak,  ezawa2018minimal, khalaf2018higher, matsugatani2018connecting, lin2018topological, dwivedi2018majorana, langbehn2017reflection, parameswaran2017topological, wang2018high, matsugatani2018connecting, trifunovic2019higher, Tiwari_2019, Tianhe_2019, You_2019, Roy_2019,ahn-yang-2020,roy-2020,hu-2020, zhang2019higherorder, zhang2020kitaev, vu2020timereversalinvariant}. 
%Within the nomenclature of higher-order topological phases, a model 
A system that supports non-trivial states on codimension-$q$ corners belong to a $q^{\text{th}}$-order topological phases. Yet another subclass of topological phases, the so-called ``boundary-obstructed topological phases"\cite{khalaf_2019,hu-2020} also host states localized on higher-codimension boundaries. However such states are not protected by the bulk energy gap. Instead as the name suggests, they are protected by the boundary energy gap (or relatedly the bulk Wannier band gap). In the literature, such a topological phase is also referred to as an ``extrinsic" higher-order topological phase~\cite{geier-2018}, in distinction with the ``intrinsic" ones protected by a bulk energy gap. {The two phases are closely related, and their connection has been studied in Ref.~\onlinecite{khalaf_2019} in the context of topological insulators, in which spatial symmetries play an important role .  In the present work we focus on topological superconductors, in which, as we shall see, the role of spatial symmetries is rather different. Specifically, we study a class of 2D superconductors that can be either second-order topological superconductors (HOTSC$_2$) or boundary obstructed topological superconductors (BOTSC$_2$) depending on the discrete rotational symmetry of the system.}

\medskip \noindent The topological properties of BdG Hamiltonians %superconductors 
in the weak pairing limit can often be understood simply and efficiently in terms of the Fermi surface properties of the normal state Hamiltonian and are independent of the details of the electronic structure away from the Fermi surface. This is not unexpected since Cooper-pairing is indeed dominant only in the neighbourhood of the Fermi surface. In 1D a gapless fermionic system with two Fermi points subject to $p$-wave pairing is a topological superconductor with Majorana zero modes at its ends \cite{Kitaev_2001}. Similarly, in 2D a system with a Fermi surface that encloses the $\Gamma$-point in the Brillouin zone, subject to chiral ${p}+i{p}$ pairing is a topological superconductor that hosts chiral Majorana modes at its edge \cite{Read_Green, PhysRevLett.86.268, RevModPhys.69.645, Beenakker, RevModPhys.83.1057, Schnyder_2008}. Similar low energy criteria have been propose for time-reversal invariant superconductors\cite{Qi_Hughes_Zhang_2010} in various dimensions. All of the above results share the remarkable feature that the topological bulk-boundary correspondence is completely contingent on the low-energy description of the normal state.  Apart from contributing towards a  clear theoretical understanding, such criteria are useful from the perspective of materials search. It is therefore highly desirable to formulate higher-order and boundary-obstructed topology in the context of superconductors in terms of similar low-energy criteria.
%\begin{table}
%\centering
% \begin{tabular}{|| p{2.5cm}| p{2.5cm} | p{2.5cm}||} 
% \hline
% \begin{center} Model \end{center} 
%  &  \begin{center} With $\ms{C}_{4}$ \end{center}   & \begin{center} With $\ms{C}_{2}$  \end{center} \\ 
% \hline\hline
% \begin{center}
%With PH
%\end{center}
%& %\rule{0pt}{4.6ex}
%\begin{center}
%HOTSC$_{2}$; \\
%%\newline with 
%corner Majorana 
%%{\rule[-3.2ex]{0pt}{0pt}} 
%\end{center}
%& 
%\begin{center}
%BOTSC$_2$; \\
%%\newline with 
%corner Majorana 
%\end{center}
%\\ 
%  \hline
% \begin{center}
% Without PH% {\rule[-3.2ex]{3pt}{0pt}}
%  \end{center} 
% &
% \begin{center}
% HOTI$_{2}$; \\
% %\rule{0pt}{4.6ex}
%% \newline with 
% filling anomaly
%% {\rule[-3.2ex]{0pt}{0pt}}
% \end{center}
%  & 
%  \begin{center}
%  Trivial
%  \end{center}
%     \\ 
%  \hline
%\end{tabular}
%\caption{A BdG superconductor with four Dirac points in the normal state Hamiltonian and an odd-parity pairing term (Eq.~\eqref{eq:model_general}) hosts Majorana zero mode corner states. When enriched by spatial $\ms{C}_4$ rotation symmetry, such a model is a second order topological superconductor (HOTSC$_2$). When the $\ms{C}_{4}$ symmetry is reduced to a $\ms{C}_{2}$ subgroup, the model continues to host Majorana zero modes as a boundary-obstructed topological superconductor.  We contrast these cases with the analogous cases without particle-hole, i.e treating the BdG Hamiltonian as an insulator wherein the phase with and without $\ms{C}_{4}$ symmetry is a second-order topological insulator (HOTI$_{2}$) and a trivial insulator respectively.  }
%\label{main_result}
%\end{table}
\begin{table}
\centering
 \begin{tabular}{|| p{2.5cm}| p{2.5cm} | p{2.5cm}||} 
 \hline
 %\begin{center} 
  \qquad Model {\rule[4.2ex]{0pt}{0pt}} 
 %\end{center} 
  & 
   %\begin{center} 
   \qquad With $\ms{C}_{4}$ 
   %\end{center}   
   & %\begin{center} 
   \qquad With $\ms{C}_{2}$  {\rule[-2.2ex]{0pt}{0pt}}%\end{center}
    \\ 
 \hline\hline
 %\begin{center}
    \parbox[c]{\hsize}{ With PH {\rule[4.2ex]{0pt}{0pt}}}
%\end{center}
& 
\qquad HOTSC$_{2}$; 
& 
\qquad BOTSC$_2$;
\\ 
 &\  corner Majorana & \ corner Majorana %{\rule[-0.2ex]{0pt}{0pt}} 
 \\
  \hline
  \parbox[c]{\hsize}{ Without PH {\rule[4.2ex]{0pt}{0pt}}}
& 
\qquad \ HOTI$_{2}$; %\\
& 
\qquad \ Trivial
\\ 
 &\ \ filling anomaly %{\rule[-2.2ex]{0pt}{0pt}} 
 & no filling anomaly  \\
  \hline
\end{tabular}
\caption{A BdG superconductor with four Dirac points in the normal state Hamiltonian and an odd-parity pairing term (Eq.~\eqref{eq:model_general}) hosts Majorana zero mode corner states. When enriched by spatial $\ms{C}_4$ rotation symmetry, such a model is a second order topological superconductor (HOTSC$_2$). When the $\ms{C}_{4}$ symmetry is reduced to a $\ms{C}_{2}$ subgroup, the model continues to host Majorana zero modes as a boundary-obstructed topological superconductor.  We contrast these cases with the analogous cases without particle-hole symmetry, i.e. by treating the BdG Hamiltonian as an insulator wherein the phase with and without $\ms{C}_{4}$ symmetry is a second-order topological insulator (HOTI$_{2}$) and a trivial insulator respectively.}
\label{main_result}
\end{table}

\medskip \noindent Toward this end, a low energy criterion %criteria in this spirit
 was proposed in Ref.~[\onlinecite{lin-wang-hughes-2018}] for a 
 topological superconductors with corner Majorana modes,
% HOTSC$_2$. 
%The criteria
which stated that a 2d HOTSC$_2$ can be realized in a doped two-band Dirac semimetal with four Dirac points in the presence of finite range attractive interactions.  With a finite density of states, i.e for a nonzero chemical potential $\mu$, the leading pairing instability is toward a ${p}+i{p}$ order. 
A ``minimal model" of such a state is given by the following Hamiltonian (see also Ref.~[\onlinecite{huang-2019}]) 
\begin{align}
H =& \int d\bm k \left\{ c^\dagger(\bm k) (t\cos k_x \sigma_x + t\cos k_y \sigma_z -\mu ) c(\bm k)\right. \nonumber \\
 & \left.+c^\dagger(\bm k) (\Delta \sin k_x + i\Delta \sin k_y) c^{\dagger}(\bm k) + \text{h.c} \right\},
\label{eq:0}
\end{align}
in which the normal state has four Dirac points at $(\pm \pi/2, \pm \pi/2)$. It has been argued that the zero modes remain robust upon various small deformations.

\medskip \noindent While Ref.~[\onlinecite{lin-wang-hughes-2018}] demonstrated the existence of four Majorana zero modes at the corners of a finite sample, it %was unclear
remained to be elucidated
 what {its topological classification is}, and what \emph{topological invariant} characterizes its nontrivial topology. One well-known topological invariant for second-order topological insulator, both intrinsic and boundary-obstructed (extrinsic), with mirror reflection symmetries is the {\it{nested Wilson loop}}\cite{ benalcazar2017quantized, benalcazar2017electric}, defined as the polarization of a Wannier band --- a band formed by orthonormal filled states that are extended in one direction but exponentially localized in the orthogonal direction. In the presence of mirror reflection symmetries the nested Wilson loop is quantized to be $0$ or $1/2$, and the second-order topology is captured by two nested Wilson loops along $x$ and $y$ directions. However, if $\mu\neq 0$, the Hamiltonian in Eq.~\eqref{eq:0} breaks mirror symmetry, and the nested Wilson loop is unquantized. 
 Further attempts to quantize the nested Wilson loop using particle-hole symmetry are also unsuccessful as explained in the main text.  {Importantly, in sharp contrast with the case of regular Wilson loop (polarization), the particle-hole symmetry does \emph{not} impose quantization conditions on nested Wilson loops.}
Consequently, {we argue} that the nested Wilson loop does not generally provide a topological invariant for {BdG Hamiltonians without mirror symmetires}.

\medskip \noindent We note that recently %there has been 
an exhaustive classification scheme of BdG Hamiltonians with additional spatial symmetries \cite{Ono_2019, Nastya_2020, geier2019}  has been developed based on organizing the BdG as well as normal state bands according to their symmetry eigenvalues at high symmetry points in the Brillouin zone. %While such a symmetry-indicator based approach is very useful and computationally efficient it does require knowledge of the full electronic and BdG band structure. 
{Within this approach, we prove that our BdG, when augmented with a $\ms{C}_4$ rotation symmetry, realizes an intrinsic HOTSC$_2$ phase. However, such an ancillary $\ms{C}_4$ symmetry is rather peculiar and artificial. In particular its corresponding operator satisfies $\widehat{\ms{C}}_4^4=-1$, and its eigenvalues are half-integers. In the presence of a lower but more natrual $\ms{C_2}$ symmetry, we find that our system is trivial in terms of intrinsic higher-order topology. Therefore this approach does not capture the BOTSC$_2$ phase in the absence of $\ms{C}_4$ symmetry.}

\medskip \noindent  {In this paper, in addition to the $\bm k$-space bulk approach described above,  we provide an alternative real-space \emph{boundary} approach.} We view corner states of {a finite-size Hamiltonian} as topological defects {of a space-filling} Hamiltonian. Since a topological defect, by definition, {being only dependent on the topological winding number}, is insensitive to the details of the band structure away from the Fermi-level we can directly work with a low-energy description of the model.  
{Further, since a topological defect is locally insensitive of the bulk rotational symmetry,} this real space approach naturally lends itself to analyzing and proving the `Dirac+${p}+i{p}$' low energy criterion for \emph{both} HOTSC$_2$'s and BOTSC$_2$'s. In order to establish this criterion we carefully study a general BdG Hamiltonian with four Dirac points in the normal state subject to odd parity pairing. The Dirac points of the normal state can be protected by a chiral symmetry, or a product of time reversal and inversion symmetry, but  both these symmetries are broken in the superconducting state. For our purposes we simply take the Dirac points as an input. In addition to particle hole symmetry such a model may have certain spatial symmetries. %A natural choice is $\ms{C}_{4}$-rotation symmetry which enforces that the four Dirac points are related to one another via the symmetry action. 
We show that depending on whether or not one imposes the additional $\ms{C}_{4}$-rotation symmetry on the model, the Majorana modes can be protected by either the bulk or the edge energy gap implying %genuine
{intrinsic} higher-order or boundary-obstructed topology respectively. The different cases with and without $\ms{C}_{4}$ symmetry are summarized in Table~\ref{main_result}. {To emphasize the role of particle-hole symmetry of the BdG Hamiltonian, we list  in Table~\ref{main_result} the topological classification of the system had we interpreted it as an insulator without particle-hole symmetry.}

\medskip \noindent The rest of the paper is organized as follows. In Sec.~\ref{Sec:model}, we introduce our model and describe its various symmetries. In Sec.~\ref{sec:sym_ind}, we study our model, enriched by $\ms{C}_{4}$-rotational symmetry on a $\ms{C}_{4}$ symmetric open geometry and show that it hosts corner Majorana modes. We first review a well established symmetry-based indicator approach in Sec.~\ref{Sec:k_space_approach} to establish the existence of corner Majoranas before moving onto a real space approach in Sec.~\ref{Sec:real_space_approach}. In Sec.~\ref{Sec:no_rotn} , we relax the rotation symmetry constraint and show that the model is non-trivial in the sense of boundary-obstructed topology. In Sec.~\ref{Sec:conclude}, we conclude with a summary and some further directions.
 
\section{Model}
\label{Sec:model}
\noindent Consider {a generic Bogoliubov-de-Gennes (BdG)} Hamiltonian described by %{(YW: As is customary, I flipped the sign of chemical potential $\mu$. Please make sure it stays consistent.)}
\begin{align}
\mathcal H(\bs{k})=\sum_{i=1,2}\left[f_{i}(\bs{k})\Gamma^{i}+ g_{i}(\bs{k})\Gamma^{i+2}\right]-\mu \Gamma^{34},
\label{eq:model_general}
\end{align}
{where $f_{i}(\bs{k})$ and $g_{i}(\bs{k})$ are even and odd functions respectively.} We use the convention $\Gamma^{1}=X^{13}$, $\Gamma^{2}= X^{33}$, $\Gamma^{3}=X^{10}$,  $\Gamma^{4}=X^{20}$ and $\Gamma^{5}=\Gamma^{1}\Gamma^{2}\Gamma^{3}\Gamma^{4}$ where $X^{\mu\nu}:=\tau^{\mu}\otimes \sigma^{\nu}$, such that $\sigma^{\mu}$ acts within the subspace of the normal-state bands while $\tau^{\nu}$  act on the Nambu space indices. {The $f_{1,2}$ terms correspond to the normal state dispersion, with  Dirac points at the common zeros of $f_{1,2}$. Constrained by periodicity and parity of $f_{1,2}$, the number of Dirac points are necessarily multiples of four. In this work, we assume that there are four such Dirac points at $\pm \bm{k}_F$ and $\pm \bm{k}_F'$. }The chemical potential term is proportional to $\Gamma^{34}:=-i\Gamma^3\Gamma^{4}$. {The pairing terms $ g_{1,2}$ are of odd parity, which in the simplest case corresponds to a $p+ip$ symmetry.
It has been shown~\cite{lin-wang-hughes-2018} that such a generic BdG Hamiltonian is the superconducting ground state of a two-band Dirac semimetal with an finite-range attractive interaction. For the rest of this paper we will take \eqref{eq:model_general} as input and analyze its topology.}

 %We require that
\medskip \noindent {By construction the BdG} Hamiltonian is particle-hole symmetric, such that
\begin{align}
\ms P \mathcal H(\bs{k}) \ms P^{-1}=-\mathcal H(-\bs{k}),
\end{align} 
where $\ms P=X^{10}\ms{K}$  and $\ms{K}$ implements complex conjugation. %This constrains the functions $f_{i}(\bs{k})$ and $g_{i}(\bs{k})$ to be even and odd respectively under the inversion transformation $\bs{k}\to -\bs{k}$. 
Furthermore if $\mu=0$, the model has an additional chiral or sublattice symmetry generated by $\ms S=\Gamma^{5}$ such that 
\begin{align}
\ms S\mathcal H(\bs{k})\ms S^{-1}=-\mathcal H(k)
\end{align}
%Additionally one may define a time reversal symmetry via $\ms T=\ms P\ms S$ which squares to +1. 
Therefore the model Eq.~\eqref{eq:model_general} with $\mu\neq 0$ belongs to the AZ class D while for $\mu=0$, it belongs to class BDI. %For the present discussion, we impose that
{To analyze its higher-order topology it will be instructive to augment}
 the model with an additional $\ms{C}_{4}$ rotation symmetry generated by %{(YW: We need to fix the $i$ factors below)}
\begin{align}
\widehat{\ms{C}}_{4}=\frac{\Gamma^{15}+\Gamma^{52}}{\sqrt{2}}\exp\left\{-\frac{i\pi \Gamma^{34}}{4}\right\},
\label{eq:C4_defn}
\end{align}
where $\Gamma^{5}=\prod_{i=1}^{4}\Gamma^{i}$. The rotation symmetry acts on the $\Gamma$-matrices as
\begin{align}
\widehat{\ms{C}}_{4}: 
\begin{bmatrix}
\Gamma^{1} \\
\Gamma^{2} \\
\Gamma^{3} \\
\Gamma^{4} \\
\Gamma^{5} 
\end{bmatrix}
\mapsto 
\widehat{\ms{C}}_{4}
\begin{bmatrix}
\Gamma^{1} \\
\Gamma^{2} \\
\Gamma^{3} \\
\Gamma^{4} \\
\Gamma^{5} 
\end{bmatrix}
\left(\widehat{\ms{C}}_{4}\right)^{\dagger}
=\begin{bmatrix}
\Gamma^{2} \\
\Gamma^{1} \\
\Gamma^{4} \\
-\Gamma^{3} \\
-\Gamma^{5} 
\end{bmatrix}.
\label{eq:C4_action}
\end{align}
The rotational symmetry defined in Eq.~\eqref{eq:C4_defn} corresponds to a double group representation as can be seen explicitly from the fact that $\widehat{\ms{C}}_{4}^{4}=-1$. %{(YW: do we need the following sentence?)} We note that the chiral symmetry anticommutes with the rotation symmetry while it commutes with the particle-hole symmetry.
Invariance of the Hamiltonian under such a rotational symmetry action further imposes the following constraints on the functions $f_{i}(\bs{k})$ and $g_{i}(\bs{k})$ in addition to the one imposed by particle-hole symmetry:
\begin{align}
f_{1,2}(\ms{C}_4\triangleright \bs{k})=&\;f_{2,1}(\bs{k}), \nonumber \\
g_{1,2}(\ms{C}_{4}\triangleright \bs{k})=&\; \pm g_{2,1}(\bs{k}). %\nonumber \\
%g_{2}(\ms{C}_{4}\triangleright \bs{k})=&\; -g_{1}(\bs{k}).
\label{eq:fg_C4_properties}
\end{align}
%We note that the constraints imposed by $\ms{C}_4$ rotation and by particle hole symmetry are compatible as they should be. 
{As we mentioned $f_{1,2}$ each have a contour of zeros that intersects at four isolated Dirac points; these four Dirac points are related by the $\ms{C}_4$ symmetry.}
A simple example of a Hamiltonian invariant such a set of symmetries is {precisely Eq.~\eqref{eq:0}. In BdG form, we have}
\begin{align}
\mathcal H(\bs{k})=\sum_{i=1,2}\left[t\cos(k_{i})\Gamma^{i}+\Delta \sin(k_i)\Gamma^{i+2}\right]-\mu \Gamma^{34}.
\label{eq:simple_form}
\end{align}
%In what follows, we consider the general Hamiltonian in Eq.~\eqref{eq:model_general} instead of the specific form in Eq.~\eqref{eq:model_general}. 
%In addition to the above mentioned symmetries, we assume that the normal state described by the Hamiltonian,
%\begin{align}
%h(\bs{k})=\sum_{i=1}^{2}f_{i}(\bs{k})\sigma^{i}-\mu \mathbb I_2,
%\end{align}
%contains four Dirac points located at points in the Brillouin zone related by $\ms{C}_{4}$ symmetry. In other words $f_{1,2}$ each have a contour of zeros that intersects at four isolated points. 

\medskip \noindent {In Sec.~\ref{Sec:no_rotn}, we will relax the $\ms{C}_4$ symmetry and analyze the fate of the second-order topology.}

\section{second-order topology from the bulk protected by rotational symmetry}
\label{sec:sym_ind}
In this section we analyze the $\ms{C}_{4}$ symmetric BdG Hamiltonian described in Eq.~\eqref{eq:model_general} and show that it is in a HOTSC$_2$ phase. We adopt two complimentary approaches to diagnose the second-order topology in such systems. Firstly, in Sec.~\ref{Sec:k_space_approach}, we use a symmetry-indicator based approach\cite{Fu_TCI_2011, Khalaf_2018, Ono_2019, Nastya_2020, geier2019} to show that when treated as a $\ms{C}_4$-symmetric band insulator, the model in Eq.~\eqref{eq:model_general} is in the obstructed atomic limit which exhibits a filling anomaly and therefore hosts corner states on four $\ms{C}_4$ related corners. Furthermore particle-hole symmetry requires that the corner states are Majorana zero modes. The results/methodology in Sec.~\ref{Sec:k_space_approach} {closely follow recent works \cite{Khalaf_2018, Ono_2019, Nastya_2020, geier2019}.} In Sec.~\ref{Sec:real_space_approach}, we adopt a real space approach to show that the $\ms{C}_4$ symmetry pins topological defects at the four corners of the aforementioned spatial geometry. Then second order topology can be demonstrated using an index theorem in conjunction with the fact that topological defects in Altland Zirnbauer class D and BDI host $\mathbb Z_{2}$ and $\mathbb Z$ classified  Majorana zero modes\cite{Teo_Kane}.  
\subsection{Momentum space approach from symmetry indicators}
\label{Sec:k_space_approach}
Let us consider our model in Eq.~\eqref{eq:model_general} as a band insulator and for the moment ignore particle-hole symmetry. The model has two occupied and two unoccupied bands. Our strategy is to first analyze the $\ms{C}_4$ eigenvalues of the occupied bands at the high symmetry points, from which we deduce the positions of the Wannier centers that induce these bands {in the ``atomic limit". We note that a proper definition of the atomic limit for superconductors has also be addressed~\cite{geier2019,neupert-fischer-2020,schindler-2020}, but for our purposes we will simply treat Eq.~\eqref{eq:model_general} as an atomic insulator.}
 It is known that the rotation symmetry indices can only determine the Chern number of the system mod four \cite{Fang_2012}. Our model with vanishing chemical potential, has an additional chiral symmetry which ensures vanishing Chern number and therefore Wannier representability. Furthermore, for small enough chemical potential that does not close the gap, which is what we assume, the model remains Wannier representable.   

\medskip \noindent Now that we know there is no Wannier obstructions, we can proceed in calculating the $\ms{C}_4$ indices.  Since the little groups at $\Gamma,M$ points and $X, X^\prime$ 
points contain the $\ms{C}_4$ and $\ms{C}_2$ symmetry operators respectively, the corresponding symmetry eigenvalues for the occupied bands are topological indices, in that they cannot be changed without a bulk gap closing. Further, we note that the ${p}+i{p}$ order parameter $g_{i}(\bs{k})$ vanishes at these high-symmetry points due to Eq.~\eqref{eq:fg_C4_properties}, which means the order parameter does not directly affect the eigenvalues at these points. However, this does not imply that these terms do not play any role in the topology of the system, they have two important effects, (i) They affect the form of the $\ms{C}_4$ operator, indeed these terms are responsible for the $\ms{C}_4^4 = -1$ property of the operator and (ii) without these terms the system is gapless which renders the eigenvalues of the $\ms{C}_{4,2}$ operators at the high-symmetry points meaningless. The $\ms{C}_4$ symmetry also imposes that $f_{1}(0,0)=f_{2}(0,0)=:f_{\Gamma}$, $f_{1}(\pi,\pi)=f_{2}(\pi,\pi)=:f_{{M}}$ and $f_{1,2}(\pi,0)=f_{2,1}(0,\pi)=:f_{X,X'}$. The Hamiltonian \eqref{eq:model_general} takes the following form at the high symmetry points
\begin{align}
\mathcal H_{\Gamma,{M}}=&\;f_{\Gamma,{M}}\left[\Gamma^{1}+\Gamma^{2}\right] \nonumber \\
\mathcal H_{{X},{X}'}=&\;f_{{X},{X}'}\Gamma^{1}+f_{{X}',{X}}\Gamma^{2}
\end{align}
 We list the eigenvales for the high symmetry points in table \ref{c_4_eigenvalues}, from which it can be seen that the eigenvalues only depend on $\text{sgn}(f_{\Gamma})$ and $\text{sgn}(f_{M})$. There are four possibilities, corresponding to distinct configurations of atomic orbitals from which the filled bands can be induced. These have been summarized in Table.~\ref{Tab:Wannier_conf}. Notably, depending on whether $\text{sgn}(f_{\Gamma})\text{sgn}(f_{M})=+1$ or $-1$, the bands can be induced from a pair of atomic orbitals localized at the Wyckoff position $1\ms{a}$ or $1\ms{b}$, i.e at $\bs{r}=(0,0)$ or $\bs{r}=(1/2,1/2)$. These two cases correspond to an unobstructed and obstructed atomic limit respectively, {as we shall see below.}
  \begin{center}
\begin{table}
 \begin{tabular}{|| c | c ||} 
 \hline
  $\bm k_*$ &  \quad $\ms{C}_{4,2}$ eigenvalues of $\mathcal{H}(k_*)$  \ \ \\ [0.5ex] 
 \hline\hline
\  \ $\Gamma=(0,0)$ \ \   & \quad $\{-\sgn(f_{\Gamma})\ e^{\frac{i\pi}{4}},\ \sgn(f_{\Gamma})\  e^{-\frac{i\pi}{4}}\}$ \quad \quad \\ 
 \hline
 $X'=(0,\pi)$ & $\{e^{\frac{i\pi}{2}}, \ e^{\frac{-i\pi}{2}}\}$  \\
 \hline
 $X=(\pi,0)$ & $\{e^{\frac{i\pi}{2}},\ e^{\frac{-i\pi}{2}}\} $  \\
 \hline
 $M=(\pi,\pi)$ & $\{-\sgn(f_{M})\ e^{\frac{i\pi}{4}}, \  \sgn(f_{M})\  e^{-\frac{i\pi}{4}}\}$ \\
 \hline
\end{tabular}
\caption{The eigenvalues of the $\ms{C}_{4}$ operators at the $\Gamma$ and $M$ points as well as the $\ms{C}_2$ eigenvalues at the $X$ and $X'$ points in the Brillouin zone. }
\label{c_4_eigenvalues}
\end{table}
\end{center}

\begin{figure}[bt]
\centering
\includegraphics[scale=0.45]{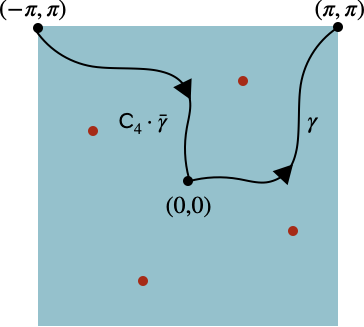}
\\
\caption{
A closed loop built from composing two $\ms{C}_4$ rotation related path segments $\gamma$ and $\ms{C}_4\cdot \gamma$ enclose a single Dirac point of the normal state Hamiltonian $\mathcal H_{\ms{nor}}(\bs{k})$ (Eq.~\eqref{eq:normal_state}). }
\label{C4_path}
\end{figure}

 \begin{center}
\begin{table}
 \begin{tabular}{|| c | c |c || } 
 \hline
  \quad $\text{sgn}(f_{\Gamma})$ \quad  &  \quad $\text{sgn}(f_{M})$  \quad  
  & \quad  \text{orbitals and Wyckoff position} \quad  \\ [0.5ex] 
 \hline\hline
\  \ $+$ \ \   & \ \ $+$ \ & $j=7/2, j=5/2$ @ $\bs{r}=(0,0)$ \ \\ 
 \hline
 \  \ $+$ \ \   & \ \ $-$ \ & $j=7/2, j=5/2$ @ $\bs{r}=({1}/{2},{1}/{2})$ \ \\ 
 \hline
 \  \ $-$ \ \   & \ \ $+$ \ & $j=3/2, j=1/2$ @ $\bs{r}=({1}/{2},{1}/{2})$ \ \\ 
 \hline
 \  \ $-$ \ \   & \ \ $-$ \ & $j=3/2, j=1/2$ @ $\bs{r}=(0,0)$ \ \\ 
 \hline
\end{tabular}
\caption{The sign of $f_{\Gamma}$ and $f_{M}$ completely determine the Wannier representation of the occupied bands. If $\text{sgn}(f_{\Gamma})\text{sgn}(f_{M})=-1$, the insulator is in an obstructed atomic limit.} %The format $j$-orbital mean an orbital with a total angular momentum $j$, so for example $(3\pm 1/2)$-orbitals means the existence of two orbitals with $j=2.5$ and $3.5$.}
\label{Tab:Wannier_conf}
\end{table}
\end{center}
Next, we confirm that our model \eqref{eq:model_general} is indeed in an obstructed atomic limit by showing that the condition $\text{sgn}(f_{\Gamma})\text{sgn}(f_{M})=-1$ follows from the fact that the normal state Hamiltonian in Eq.~\eqref{eq:model_general} contains a single Dirac cone per Brillouin zone quadrant.  The normal state Hamiltonian takes the form
\begin{align}
\mathcal H_{\ms{nor}}(\bs{k})=&\;f_{1}(\bs{k})\sigma^{x}+f_{2}(\bs{k})\sigma^{z} =: ||f||\hat{\bs{n}}\cdot\bs{\sigma},
\label{eq:normal_state}
\end{align}
where, $||f(\bs{k})||$ and $\hat{\bs{n}}$ are the norm and unit vector corresponding to $\bs{f}:=(f_1(\bs{k}),f_2(\bs{k}))$. The $\ms{C}_4$ transformation acts within the normal state via the operator
\begin{align}
\widehat{\ms{C}}^{\ms{nor}}_4:=\frac{1}{\sqrt{2}}\left(\sigma^{x}+\sigma^{z}\right),
\end{align}
as a mirror reflection about the $(1,1)$ axis in $x-z$ plane in spinor space, i.e $\widehat{\ms{C}}^{\ms{nor}}_4$, i.e $\widehat{\ms{C}}^{\ms{nor}}_4:(\sigma^{x},\sigma^{z})\to (\sigma^{z},\sigma^{x})$. The occupied state for the Hamiltonian Eq.~\eqref{eq:normal_state}, i.e $|-\hat{\bs{n}}(\bs{k})\rangle$ therefore satisfies 
\begin{align}
\widehat{\ms{C}}^{\ms{nor}}_4|\hat{\bs{n}}(\bs{k})\rangle=&\; e^{i\lambda}|\ms{M}_{(1,1)}\cdot \hat{\bs{n}}(k)\rangle \nonumber \\
=&\; e^{i\lambda} | \hat{\bs{n}}(\ms{C}_4\cdot \bs{k})
\rangle,
\end{align}
where $\ms{M}_{(1,1)}$ implements a reflection about the $(1,1)$ axis in spinor-space. It follows that the spinor at rotation invariant points $\Gamma$ and $M$ must correspond to $\hat{\bs{n}}(\bs{k}_{\Gamma,M})= \pm (1/\sqrt{2},1/\sqrt{2})$ which correspond to $\text{sgn}(f_{\Gamma,M})=\mp 1$. Now consider a closed loop constructed from two $\ms{C}_4$-related paths $\gamma$ and $\ms{C}_4\cdot\bar{\gamma}$ as illustrated in Fig.~\ref{C4_path}. Such a loop encloses a single Dirac cone and therefore the spinor wavefunction must wind an odd number of times when traversing this closed path. Let us denote the winding number around a path $\gamma$ as $\ms{N}_{\ms{w}}(\gamma)$, then
\begin{align}
\ms{N}_{\ms{w}}\left(\gamma\circ (\ms{C}_4\cdot \bar{\gamma})\right)=2\ms{N}_{\ms{w}}(\gamma)
\end{align}
where $\circ$ denotes composition of paths. {Therefore, we have $\ms{N}_{\ms{w}}(\gamma)\in \mathbb Z+1/2$.  It can readily be seen that this can be achieved only if $\hat{\bs{n}}(\bs{k}_{\Gamma})=-\hat{\bs{n}}(\bs{k}_{M})$. }%, then $\ms{N}_{\ms{w}}(\gamma)\in \mathbb Z+1/2$ whereas if $\hat{\bs{n}}(\bs{k}_{\Gamma})=\hat{\bs{n}}(\bs{k}_{M})$ then $\ms{N}_{\ms{w}}(\gamma)\in \mathbb Z$. 
This concludes the proof that $\text{sgn}(f_{\Gamma})\text{sgn}(f_{M})=-1$. {We will be again using this relation in Sec.~\ref{Sec:real_space_approach}.}

\begin{figure}[t]
    \centering
    \subfigure[ ]{
    \begin{tikzpicture}[scale = 0.95]
        \draw[black] (0,0) grid(4,4);

        \foreach \i in {0,...,3} 
        \foreach \j in {0,...,3} 
        {
        \draw[black] (\i+0.45,\j+0.45) -- (\i+0.55,\j+0.55);
        \draw[black] (\i+0.45,\j+0.55) -- (\i+0.55,\j+0.45);
        }

        \foreach \i in {1,...,3} 
        \foreach \j in {1,...,3} 
        {
        \draw[fill=blue, very thin] (\i+0.03 ,\j+0.03) circle (1.8pt);
        \draw[fill=blue, very thin] (\i-0.03 ,\j-0.03) circle (1.8pt);
        
        }

        \foreach \i in {1,...,3} 
        {
        \draw[fill=cyan, very thin] (0 ,\i ) ellipse (1.5pt and 3pt);
        \draw[fill=cyan, very thin] (4 ,\i ) circle (1.5pt and 3pt);
        \draw[fill=cyan, very thin] (\i ,0 ) circle (3pt and 1.5pt);
        \draw[fill=cyan, very thin] (\i ,4 ) circle (3pt and 1.5pt);
        }
    \end{tikzpicture}
    }
    \subfigure[ ]{
        \begin{tikzpicture}[scale = 0.95]
            \draw[black] (0,0) grid(4,4);

            \foreach \i in {0,...,3} 
            \foreach \j in {0,...,3} 
            {
            \draw[black] (\i+0.45,\j+0.45) -- (\i+0.55,\j+0.55);
            \draw[black] (\i+0.45,\j+0.55) -- (\i+0.55,\j+0.45);
            }
    
            \foreach \i in {1,...,3} 
            \foreach \j in {1,...,3} 
            {
            \draw[fill=blue, very thin] (\i+0.03 ,\j+0.03) circle (1.8pt);
            \draw[fill=blue, very thin] (\i-0.03 ,\j-0.03) circle (1.8pt);
            }
    
            \foreach \i in {1,...,3} 
            {
            \draw[fill=cyan, very thin] (0 ,\i ) ellipse (1.5pt and 3pt);
            \draw[fill=cyan, very thin] (4 ,\i ) circle (1.5pt and 3pt);
            \draw[fill=cyan, very thin] (\i ,0 ) circle (3pt and 1.5pt);
            \draw[fill=cyan, very thin] (\i ,4 ) circle (3pt and 1.5pt);
            }
            \draw[fill=red, very thin] (0 ,0 ) circle (1.5pt);
            \draw[fill=red, very thin] (4 ,0 ) circle (1.5pt);
            \draw[fill=red, very thin] (0 ,4 ) circle (1.5pt);
            \draw[fill=red, very thin] (4 ,4 ) circle (1.5pt);
        \end{tikzpicture}
        }
\caption{Depiction of the Wannier centers for a sample with boundaries. The blue circles are bulk Wannier orbitals and there are 2 orbitals per unit cell both sitting at the Wcykoff position $\bm r = (1/2,1/2)$. (The slight offsetting in the figure is just a visual aid to emphasis the existence of two centers). The cyan ellipses are boundary orbitals and they represent only one Wannier center. {When interpreted as an insulator system, filling the orbitals in a $\ms{C}_4$ symmetric manner necessarily leads to a violation of charge neutrality at the corners. For a BdG Hamiltonian this corresponds to corner Majorana zero modes.} } 
\label{fig:filling_anomaly}
\end{figure}
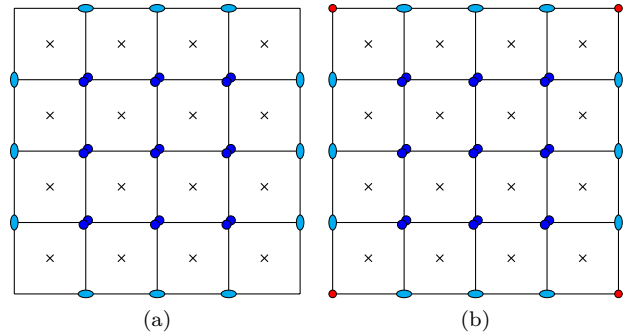

\medskip \noindent  Interpreting the BdG Hamiltonian as a $\ms{C}_4$ symmetric \emph{insulator} with open boundary conditions, it can be shown that for the phase with Wannier centers at (1/2,1/2) there is no way to satisfy both $\ms{C}_4$ symmetry and neutrality. This phenomenon, known as the filling anomaly, necessitates corner states as long as the spatial rotation symmetry is unbroken~\cite{Taylor_Filling}. It is important to check that the system has no polarization, otherwise the edges will be gapless and it would not make sense to talk about corner modes. That the polarization is zero can be checked by simple counting of the charges as shown in Fig.~\ref{fig:filling_anomaly}. Configuration (a) in Fig.~\ref{fig:filling_anomaly} has the minimum amount of electrons to achieve neutrality of bulk and the edges, however it is two electrons short for overall system neutrality. Since the system is $C_4$ symmetric we will have $+e/2$ charge localized on each corner. Two electrons cannot be added to the system in a $C_4$ symmetric manner. The other closest configuration to neutrality is shown in Fig \ref{fig:filling_anomaly} (b). In this case the system has two electrons in excess than that needed for neutrality, and hence $-e/2$ charge localized on each quadrant.

\medskip \noindent {Finally, recalling that the system under study is a superconductor with a particle-hole symmetric BdG Hamiltonian, the filling anomaly is manifested as corner Majorana zero modes. }

\subsection{{Real space approach from defect classification }}
\label{Sec:real_space_approach}
In this section we construct a real space topological invariant to diagnose the purported Majorana corner states {by treating the corner of a finite system as a topological defect of a \emph{space-filling} Hamiltonian.}

\medskip \noindent Consider placing the  model in Eq.~\eqref{eq:model_general} on an open $\ms{C}_{4}$ symmetric spatial geometry $M$ such that the region outside $M$ corresponds to a BdG Hamiltonian with the same form as Eq.~\eqref{eq:model_general} but where the normal state, being featureless has a vanishing fermi surface.  For concreteness, we may assume that the `outside' is described by a BdG Hamiltonian  %{(YW: I flipped the sign of $\mu$ below)}
\begin{align}
\mathcal H_{\ms{triv}}(\bs{k})=\sum_{i=1,2}\left[-f_0\Gamma^{i}+\Delta \sin(k_{i})\Gamma^{i+2}\right]-\mu \Gamma^{34},
\label{eq:model_trivial}
\end{align}
where $f_0>0$. Note that we have assumed a specific ${p}+i{p}$ form of the pairing potential for the sake of simplicity of presentation, however the analysis would not change had we chosen general functions $g_{i}(\bs{k})$ that satisfy the symmetry constraints in Eq.~\eqref{eq:fg_C4_properties}. In going from inside to outside, the sign of ${f}_{i}$ must change at exactly one of the high symmetry points $\Gamma$ or $M$, since $\text{sgn}(f_{\Gamma})\text{sgn}(f_{M})=-1${, as we proved earlier}. Therefore the boundary physics is determined entirely by the vicinity of that particular high-symmetry point. The Hamiltonian about the two high symmetry points have the following Dirac-like forms
\begin{align}
\mathcal H(\bs{k}_{\Gamma}+\bs{q})=&\;\sum_{i}\left[f_{\Gamma}\Gamma^{i}+ \Delta q_{i}\Gamma^{i+2}\right]-\mu \Gamma^{34}, \nonumber \\
\mathcal H(\bs{k}_{M}+\bs{q})=&\;\sum_{i}\left[f_{M}\Gamma^{i}- \Delta q_{i}\Gamma^{i+2}\right]-\mu \Gamma^{34}.
\label{eq:gamma_pt_ham}
\end{align}
Both these Hamiltonians belong to class D or BDI depending on whether the chemical potential is vanishing or not as the chiral symmetry generated by $\ms S=\Gamma^{5}$ is broken by the chemical potential term. It is known that topological defects in class D and BDI host Majorana zero modes and are classified by $\mathbb Z_{2}$ and $\mathbb Z$ respectively\cite{Fukui_2010,Teo_Kane}.  Here we first explicitly compute the invariant associated to a defect (localized at a given corner of $M$) using the chiral index theorem for the case of $\mu=0$. For the case of $\mu \neq 0$, it is known that the index is simply the chiral index evaluated modulo 2  as long as the chemical potential is small enough such that it does not close the gap thereby potentially changing the topological invariant\cite{Shiozaki_2012}.

\begin{figure}[bt]
\centering
\includegraphics[scale=0.45]{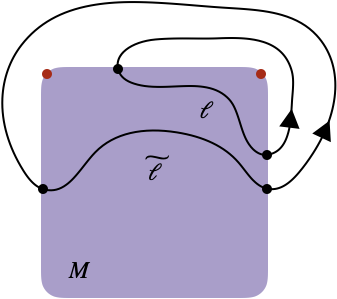}
\\
\caption{
A closed loop built from composing two $\ms{C}_4$ rotation related path segments $\gamma$ and $\ms{C}_4\cdot \gamma$ enclose a single Dirac point of the normal state Hamiltonian $\mathcal H_{\ms{nor}}(\bs{k})$ (Eq.~\eqref{eq:normal_state}). }
\label{Real_space_path}
\end{figure}

\medskip \noindent For concreteness let $f_0=f_{\Gamma}>0$, and $f_{M}<0$ in Eq.~\eqref{eq:gamma_pt_ham}. Then the low energy physics is well described by the Dirac Hamiltonian linearized about the $\Gamma$-point. In order to show the existence of a Majorana zero mode on the edge, we consider a loop (See Fig.~\ref{Real_space_path}) $\ell$ parametrized by the angular variable $\theta$, which intersects the edge $\partial M$  at points $x_{0}$ and $x_1$ which are related by a $\ms{C}_{4}$ rotation. The mass of the Dirac Hamiltonian take values $f_0$ inside, $-f_0$ outside and vary smoothly in a narrow region close to the edges. {For a general point in space there are two Dirac mass terms allowed, $\ms{m}_{1,2}(\bm{r})\Gamma^{1,2}$. Without loss of generality we assume that along the path $\ell$ the norm of the mass is constant, i.e., $\sqrt{\ms{m}_1^2+\ms{m}_2^2}=f_0:=\ms{m}$.} We parametrize the masses by a single variable $\Phi(\theta)$ such that the defect Hamiltonian takes the form of a continuum Dirac model
\begin{align}
\mathcal H(\bs{q},\theta)=\sum_{i}\Delta q_{i}\Gamma^{i+2}
+
\ms{m}\left[\cos (\Phi)\Gamma^{1}+ 
\sin (\Phi)\Gamma^{2}
\right]
\label{eq:BDI_defect_Ham}
\end{align}
{Interestingly, we note that this defect Hamiltonian is dual to that for the celebrated Fu-Kane superconductor~\cite{fu-kane-2008}, which describes the $s$-wave pairing vortex of a Dirac fermion -- while the first term come from pairing and the second from normal state band structure in our case, it is the opposite in the Fu-Kane superconductor.}
It is well-known\cite{Teo_Kane} that a point-like mass defect in a class BDI Hamiltonian traps $\ms{N}
_{\mathsf{w}}$ majorana zero modes where $\ms{N}_{\ms{w}}$ corresponds to the winding number of the defect. 

\medskip \noindent {Consider a continuum model of the form $\mathcal H(\bs{q},\theta)=\sum_{i=1}^{4}\hat{n}_i \Gamma^{i}$ where $\bs{\hat{n}}$ is a map from $\mathbb R^{2}\times S^{1}$ to $S^{3}$. Such a Hamiltonian can be obtained  from Eq.~\eqref{eq:BDI_defect_Ham}, by simply including an overall normalization factor of $\sqrt{\Delta^{2}|\bs{q}|^2 +\ms{m}^2}$ which does not alter the winding number as it is positive definite. Furthermore by adding a $\ms{C}_4$-symmetric $|\bs{k}|\to \infty$ regularization such as $\epsilon |\bs{k}|^2 (\Gamma^{1}+\Gamma^{2})$ and taking eventually taking the $\epsilon\to 0$ limit, the field $\bs{n}$ denotes a map from $S^{3}$ to $S^{3}$. The topological invariant then is the winding number of this map, which simplifies to the winding number of $\Phi$:
\begin{align}
\ms{N}_{\ms{w}}:=&\; \frac{1}{4\pi^2} \int_{S^{3}} \epsilon^{ijkl}{\bs{n}}_{i}\mathrm{d}{\bs{n}}_j \wedge \mathrm{d}{\bs{n}}_k \wedge \mathrm{d}{\bs{n}}_l \nonumber \\
 =&\;  \frac{1}{2\pi}\oint_{\ell} \mathrm{d}\Phi.
\end{align}}
Let $\theta=0$ correspond to a point in the bulk and $\theta=\pi$ outside. The points $x_0,x_1\in \partial M$ correspond to $\theta=\alpha$ and $-\alpha$ respectively. Then it can be seen that, $\Phi(0)=\pi/4$ while, $\Phi(\pi)=5\pi/4$. Finally close to $x_0$ and $x_{1}$, $\Phi(-\alpha)=\pi/2-\Phi(\alpha)$ due to the $\ms{C}_4$ action defined in Eq.~\eqref{eq:C4_action}. Using these relations it can be shown that
\begin{align}
\ms{N}_{\text{w}}(\gamma)=&\;\frac{1}{2\pi}\oint_{0}^{2\pi} \mathrm{d}\Phi(\theta) \nonumber \\
=&\; \frac{1}{2\pi}\int_{0}^{\pi} \mathrm{d}\Phi(\theta)+\frac{1}{2\pi} \int_{\pi}^{0} \mathrm{d}\Phi(-\theta) \nonumber \\
=&\; \frac{1}{\pi}\int_{0}^{\pi}\mathrm{d}\Phi(\theta) \nonumber \\
=&\; (2n+1)
\label{corner_quantization}
\end{align}
which signals that there are an odd number of zero modes localized on the edge between $x_0$ and $x_1$. For the case of finite chemical potential $\mu$, this index is reduced modulo 2 and therefore there is a single Majorana zero mode localized between $x_0$ and $x_1$. Similarly we may consider the winding number around a loop $\widetilde{\ell}$ that intersects $\partial M$ at two points $\tilde{x}_0$ and $\tilde{x}_1$ that are related by a $\ms{C}_{2}$ rotation. The symmetry constraint imposes that $\Phi(-\theta)=\frac{\pi}{2}-\left[ \frac{\pi}{2}-\Phi(\theta)\right]=\Phi(\theta)$, therefore the winding number corresponding to such a loop evaluates to
\begin{align}
\ms{N}_{\text{w}}(\widetilde{\ell})=&\; \frac{1}{2\pi}\oint_{0}^{2\pi} \mathrm{d}\Phi(\theta) \nonumber \\
=&\; \frac{1}{2\pi}\int_{0}^{\pi} \mathrm{d}\Phi(\theta)+ \frac{1}{2\pi}\int_{\pi}^{0} \mathrm{d}\Phi(-\theta) \nonumber \\
=&\; 0.
\end{align}
Consequently, a path of the type $\widetilde{\ell}$ encloses two Majorana zero modes with opposite topological indices.

\section{Boundary obstructed topology in the absence of $\widehat{\ms{C}}_4$}
\label{Sec:no_rotn}
In this section we focus on characterization of the topology of the chiral Dirac superconductor in the  \emph{absence} of $\ms{C}_4$ rotation symmetry. The Dirac points in the normal state are generically described by the following Hamiltonian
\be
\mathcal H_{\rm nor}(\bs{k}) = f_1(\bm k) \sigma_x + f_2(\bm k)\sigma_z,
\ee
in which $f_{1,2}(\bm k)$ are even functions and the location of the Dirac points are given by their simultaneous zeros. We assume as input that there are four Dirac points at $\pm (k_{x0}, k_{y0})$ and $\pm (k_{x0}', k_{y0}')$. The Dirac points are locally (in $\bm k$-space) protected by a ``minimal" symmetry which is the product of a time-reversal $\ms{T}=\ms{K}$ symmetry and a $\widehat {\ms{C}}_2=\widehat {\ms{C}}_4^2$ rotation symmetry. For our propose, we assume that the two symmetries are separately preserved, which makes Cooper pairing energetically favorable. In the presence of ${p}+i{p}$ pairing, the time reversal symmetry is broken, and the only symmetry of the BdG Hamiltonian in consideration is a $\ms{C}_2$ rotational symmetry.

\subsection{Bulk topological invariants}
We note that just like $\ms{C}_4$ symmetry,  $\ms{C}_2$ symmetry can also potentially protect intrinsic higher-order topological phases. However, it can be confirmed by checking the symmetry-based indicators 
 that our Hamiltonian is in the \emph{trivial} phase as classified by $\ms{C}_2$ symmetry. To see this, note that the two Wannier orbitals at $\bm r= (1/2, 1/2)$ found in Sec.~\ref{sec:sym_ind} can be adiabatically moved in opposite directions, and each can merge with a Wannier orbital from the neighboring Wannier orbital at the Wyckoff position $\bm r=(0,0)$, trivializing the phase without closing the bulk energy gap. Such a process is forbidden by the $\ms{C}_4$ symmetry but allowed by the $\ms{C}_2$ symmetry.

 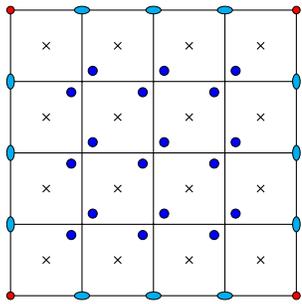
\begin{figure}[t]
    \centering
    \begin{tikzpicture}[scale = 0.95]
        \draw[black] (0,0) grid(4,4);

        \foreach \i in {0,...,3} 
        \foreach \j in {0,...,3} 
        {
        \draw[black] (\i+0.45,\j+0.45) -- (\i+0.55,\j+0.55);
        \draw[black] (\i+0.45,\j+0.55) -- (\i+0.55,\j+0.45);
        }

        \foreach \i in {1,...,3} 
        \foreach \j in {1,...,3} 
        {
        \draw[fill=blue, very thin] (\i+0.15 ,\j+0.15) circle (1.8pt);
        \draw[fill=blue, very thin] (\i-0.15 ,\j-0.15) circle (1.8pt);
        }

        \foreach \i in {1,...,3} 
        {
        \draw[fill=cyan, very thin] (0 ,\i ) ellipse (1.5pt and 3pt);
        \draw[fill=cyan, very thin] (4 ,\i ) circle (1.5pt and 3pt);
        \draw[fill=cyan, very thin] (\i ,0 ) circle (3pt and 1.5pt);
        \draw[fill=cyan, very thin] (\i ,4 ) circle (3pt and 1.5pt);
        }
        \draw[fill=red, very thin] (0 ,0 ) circle (1.5pt);
        \draw[fill=red, very thin] (4 ,0 ) circle (1.5pt);
        \draw[fill=red, very thin] (0 ,4 ) circle (1.5pt);
        \draw[fill=red, very thin] (4 ,4 ) circle (1.5pt);
    \end{tikzpicture}
\caption{A snapshot of the Wannier orbitals during a $\ms{C}_2$-{symmetric} process that connects the bulk Wannier centers from the Wyckoff positions $\bm r=(0,0)$ to $r=(1/2,1/2)$. As argued in the main text, the {edge} orbitals are pinned to the $\bm r=(1/2,1/2)$ because of particle-hole symmetry and $\ms{C}_2$ symmetry. This shows the importance of particle-hole symmetry for the boundary obstructed phase.} 
\label{fig:fixed_wannier_boundary}
\end{figure}

\medskip \noindent However, particle-hole symmetry ensures that the corner Majorana zero modes are stable unless there is a gap closing, in the bulk or on the boundary. We show that, in this case the second-order topology is extrinsic.~\cite{geier-2018} In our case without a topological invariant for the energy bands, the bulk energy gap does not need to close when the Majorana modes disappear. In this situation the Majorana zero modes can vanish via a gap closing at the edges only. In the terminology introduced in Ref.~\onlinecite{khalaf_2019} the corner Majorana modes are protected by a boundary obstruction. {Indeed, from the perspective of real-space Wannier orbitals, while in the bulk the two Wannier orbitals can continuously move from  $\bm r= (1/2, 1/2)$ to $\bm r= (0, 0)$, at the boundary the sole filled Wannier orbital per unit cell cannot move. This is because in order to maintain zero polarization for the completely filled bands the empty Wannier orbital (For convenience we refer to the negative (positive) energy states of the BdG Hamiltonian as filled (empty) bands.)
 would have to move in the opposite direction, which then violates on-site particle-hole symmetry. We illustrate such a situation in Fig.~\ref{fig:fixed_wannier_boundary}.}
 
\begin{figure*}[t]
    \centering
    \subfigure[]{\includegraphics[trim= 0 10 0 0, clip,scale=0.9]{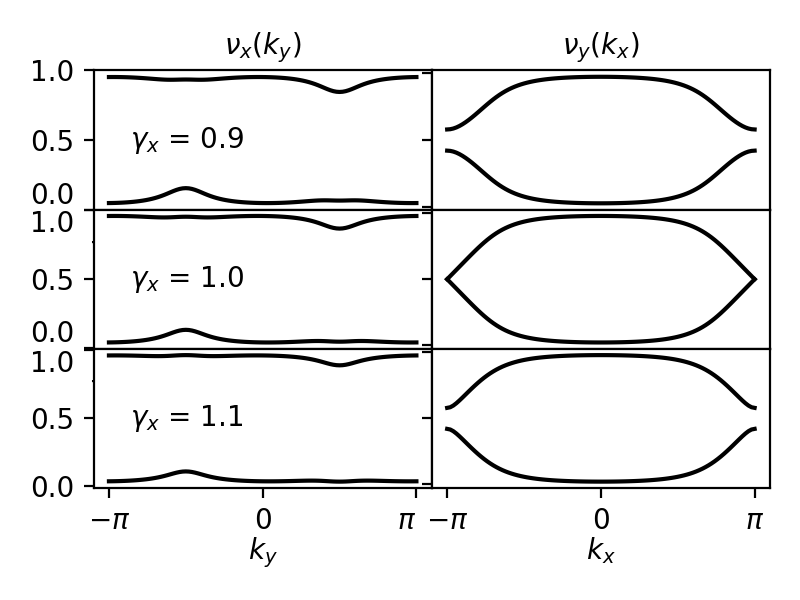}}
    \subfigure[]{\includegraphics[scale=0.85]{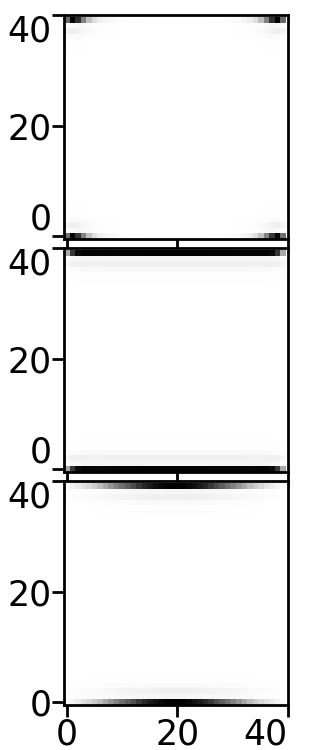}}
    \subfigure[]{\includegraphics[scale=0.9]{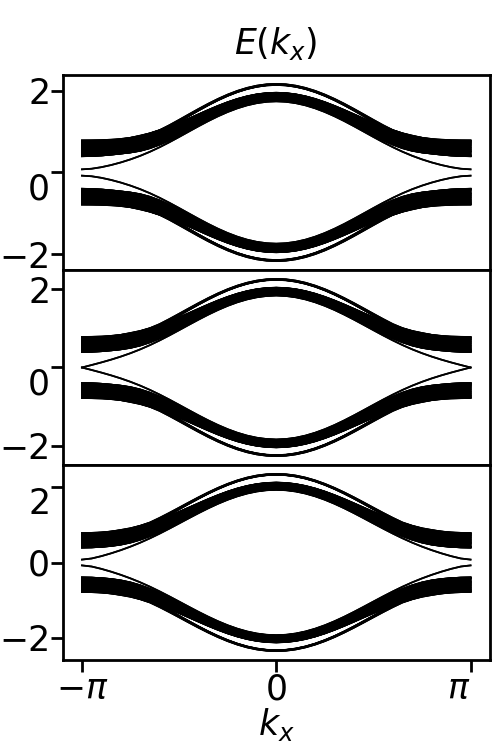}}
    \caption{Snapshots of the Wannier spectrum as $\gamma_x$ is changed. $\mu=0.2$ and $\Delta=0.4$ for all graphs. Top, middle, and bottom panels are for $\gamma_x=0.9, 1$ and $1.1$ respectively. (a) Shows the Wannier bands for Wilson loops along $x$ ($y$) on the left (right). In studying the system with open boundary conditions in both the $x$ and the $y$ directions with grid size $(40$x$40)$, (b) shows the probability densities for the 4 wavefunctions with the lowest energies. (c) Shows the spectrum of the system in a cylindrical geometry, with periodic boundary condition along the $x$ direction. The size of the system along the $y$ direction is 40. The graph shows the lowest 100 levels.}
    \label{fig:gamma_x_snapshots}
\end{figure*}

\medskip \noindent A boundary obstructed topological superconductor and a trivial superconductor with its edge wrapped by a 1d topological superconductor are similar in the sense that in both cases nontrivial topology is hosted on a 1d subsystem.
However, they can be distinguished by the fact that in the former case the edge topology ultimately comes from bulk properties. One such example is the quantized quadrupole insulator~\cite{benalcazar2017quantized, benalcazar2017electric, khalaf_2019} in which the fractional corner charge and the fractional  edge polarization comes from a fractional bulk quadrupole moment. As a result, an edge gap-closing transition is indeed captured by a bulk transition. However, such a transition happens in the bulk \emph{Wannier bands}, rather than the energy bands. The Wannier band describes the spectrum, as a function of momentum along one direction, say $k_x$, of the projected position operator (also known as the Wannier Hamiltonian for a lattice system) {$\hat\nu_y(k_x)$. Loosely speaking $\hat\nu_y(k_x)\sim \hat P_{\rm occ}(\bm k)\hat y\hat P_{\rm occ}(\bm k)$, where $\hat y$ is defined modulo 1 for a lattice system,  and $\hat P_{\rm occ}{(\bm k)}$ is the projection operator onto the filled bands. More rigorously, according to the definition in Ref.~\onlinecite{benalcazar2017quantized, benalcazar2017electric},
\be
\hat\nu_y(k_x) \equiv \frac{1}{2\pi i L_y}\log\prod_{n=0}^{L_y-1} \hat P_{\rm occ}\(k_y+\frac{2\pi n}{L_y}, k_x\),
\ee
where $L_y$ is the system size along $y$ direction. $\hat\nu_x(k_y)$ is not single valued but defined modulo 1. For definiteness we set $0\leq \hat\nu_x(k_y) < 1$.
\footnote{One subtlety is that the eigenstates of the Wannier Hamiltonian is \emph{not} a Wannier state, but rather a Bloch state with definite $k_x$ and $k_y$. Its eigenvalues, on the other hand, are independent on $k_x$ and correspond to the center position of the Wannier states. We have suppressed the $k_x$ argument in  the definition of $\hat\nu_x(k_y)$.}
}

 The eigenvalues $\{\nu_{x,\pm} (k_y)\}$ of the Wannier Hamiltonian corresponds to the center of the  \emph{hybrid Wannier states}, localized in $x$ direction and extended in $y$ direction with momentum $k_y$.
In our case, we find that the edge gap closing upon which the corner Majorana zero modes disappears is also detected by a bulk Wannier gap-closing transition. As a concrete example, we consider the following BdG Hamiltonian:
\begin{align}
H= & (\cos k_x+ \gamma_x) \sigma_x\tau_z +  \cos k_y  \sigma_z\tau_z - \mu \tau_z \nonumber\\
&+ \Delta \sin k_x \tau_x + \Delta \sin k_y \tau_y
\label{eq:example}
\end{align}
where normal state Dirac points and corner Majorana zero modes exists for $|\gamma_x|<1$ (see Fig.~\ref{fig:gamma_x_snapshots}b). At $\gamma_x=1$ there is a gap closing in the $x$-edge spectrum, through which the corner Majorana modes annihilates.  At the same point, we show in Fig.~\ref{fig:gamma_x_snapshots}a that there is a Wannier gap closing for the Wannier Hamiltonian $\hat{\nu}_{y}(k_x)$ at $k_x=0$, exactly where the edge gap closes.

\medskip \noindent One may ask if the bulk Wannier transition is captured by the change in a bulk topological invariant. In the presence of mirror symmetries, such a topological invariant is the Berry phase of the 1d Wannier band, known as the nested Wilson loop~\cite{benalcazar2017quantized, benalcazar2017electric, khalaf_2019}, which is quantized by symmetry to take the value $0$ or $1/2$. In our case, the ${p}+i{p}$ pairing order in general violates all mirror symmetries. One may naively expect that the particle-hole symmetry to quantize the nested Wilson loop. However, the two Wannier bands from the filled states with opposite eigenvalues of the Wannier Hamiltonian  $\hat\nu_y(k_x)$ are not related by particle-hole symmetry -- the two bands form a Hilbert subspace of the filled bands only, and particle-hole conjugation does not transform one to the other. Due to the $\ms{C}_2$ symmetry, the Wannier Hamiltonian satisfies
$\hat\nu_y(k_x) = - \hat{\ms{C}}_2 \hat\nu_y(-k_x) \hat{\ms{C}}_2$,
which can be thought of as a composite of chiral symmetry and 1d inversion symmetry. However, unlike the mirror symmetry, this symmetry does not host any topologically nontrivial classifications.

\medskip \noindent To circumvent this difficulty with the Wannier Hamiltonian, one  can consider the pair of Wannier bands from the filled state and the empty states with the same eigenvalue for $\hat\nu_y$ and $\hat\nu_y'\sim \hat P_{\rm emp}(\bm k)\hat y\hat P_{\rm emp}(\bm k)$, repectively ($\hat P_{\rm emp}=1-\hat P_{\rm occ}$ is the projection operator to the empty bands). They are eigenstates of the following \emph{Wannier-projected Hamiltonian} 
\begin{align}
H_{P^\pm}(k_x)=&P^\pm (k_x) H P^\pm (k_x)
, \textrm{where} \nonumber\\
P^\pm (k_x)\equiv&\frac{1\pm\hat P_{\rm occ}(\bm k)\sgn(\hat \nu_y)\pm\hat P_{\rm emp}(\bm k)\sgn(\hat \nu_y')}{2}
% \hat P_{\rm emp}(\bm k)\frac{1\pm \sgn(\hat x)}{2} \hat P_{\rm emp}(\bm k) \nonumber\\
%&+\hat P_{\rm occ}(\bm k) \frac{1\pm \sgn(\hat x)}{2}\hat P_{\rm occ}(\bm k).
\end{align}
{The Wannier-projected Hamiltonian is not to be confused with the Wannier Hamiltonian, but it shares the same eigenstates with $\nu_y(k_x)$ [$\nu_y'(k_x)$] for filled (empty) bands. When the Wannier gap closes, the eigenstates $|\nu_{y,\pm}(k_x)\rangle$ develop a singularity, which can be detected both in $\hat\nu_y(k_x)$ and $H_{P^\pm}(k_x)$.}

\medskip \noindent In the Hilbert subspace subtended by these two bands, particle-hole conjugation indeed transform one band to the other. However, since the projection operators $\hat P^\pm$ depends on $k_y$ and are thus non-onsite in the $y$ direction, the projected particle-hole symmetry operator is also non-onsite. Indeed, it is not difficult to show that the projected particle-hole operator for the Wannier projected Hamiltonian $H_{P^\pm}(k_x)$  is given by 
%\addAJ{Should one think of the spectrum of this projected Hamiltonian to be related to the spectrum of the actual Hamiltonian or the Wannier spectrum?} YW: actual Hamiltonian
\begin{align}
{\tilde{\ms{P}}}(k_x) K =& P^\pm ( k_x)\ms{P} \[P^{\pm}(- k_x)\]^* K \nonumber\\
H_{P^\pm}(k_x) =& - {\tilde{\ms{P}}}(k_x) H^*_{P^\pm}(-k_x) {\tilde{\ms{P}}}^\dagger(k_x)
\end{align}
Such a non-onsite operator does not quantize the Berry phase of the Wannier band, i.e., the nested Wilson loop, {since in the proof of Berry phase quantization by the regular particle-hole symmetry, one needs to commute the particle-hole operator with $\partial_{k}$.~\cite{qi-hughes-zhang-2008}}
Indeed, the nested Wilson loop can be computed via 
\begin{align}
    \hat\nu^{\pm x}_y(k_x)  \equiv & \frac{1}{2\pi i L_y}\log\prod_{n=0}^{L_y-1} \hat P^{\pm{x}}\(k_x, k_y +\frac{2\pi n}{L_y} \),  \nonumber  \\
    \hat P^{\pm{x}}\(\bm k\) \equiv & \frac{1 \pm \hat P_{\rm occ}(\bm k)\sgn(\hat \nu_x)}{2},
\end{align}
A direct evaluation of the nested Wilson loop $\bm P^{ \nu}(\bm k) = (\nu^{+y}_x(k_y), \nu^{+x}_y(k_x))$ shows that they are not only unquantized but also depend on their respective starting point $k_{y,x}$.
For the Hamiltonian \eqref{eq:example} at $\Delta = 0.4$, and $\mu = 0.2$, gives $\bm P^{ \nu}(\frac{\pi}{2},\frac{\pi}{2}) = (0.46, 0.5)$ for $\gamma_x = 0.9$ and $\bm P^{ \nu}(\frac{\pi}{2},\frac{\pi}{2}) = (0.96, 0.5)$ for $\gamma_x = 1.1$.

\medskip \noindent Still, the bulk Wannier transition corresponds to the change in a topological invariant. With the projected particle-hole symmetry ${\tilde{\ms{P}}}(k_x)$, the Wannier transition occurs through a gap closing at high-symmetry momenta, e.g., $k_x=0,\pi$. %\addAJ{but this particle-hole operator act on the projected Hamiltonian not the Wannier bands.} YW: they have the same eigenstates, so they have the same nested Wilson loop.
The Wannier-projected Hamiltonian $H_{P^\pm}(k_x=0,\pi)$, as two zero-dimensional subsystems, are invariant under the respective projected particle-hole symmetry ${\tilde{\ms P}}(k_x=0,\pi) \ms K$ and do each admit a $\mathbb{Z}_2$ classification~\cite{qi-hughes-zhang-2008, Ryu_2010,Ching-Kai_2016}. The Wannier transition, through which the corner Majorana modes annihilate, is thus captured by the change in one of the  $\mathbb{Z}_2$ invariants {of the Wannier-projected Hamiltonian}. 
However,  since the particle-hole symmetry $H_{P^\pm}$ at $k_x=0$ and $k_x=\pi$ are different, we cannot use  their $\mathbb{Z}_2$ invariants to construct a ``relative parity" that captures the property of the whole band $\{|\nu_{y,\pm}(k_x)\rangle\}$. As a result, these  $\mathbb{Z}_2$  invariants do not capture the presence or absence of corner Majorana modes -- it only detects the change in their existence. In the next section we directly show the existence of the corner Majorana modes from low-energy properties of the model through the defect classification approach.

\subsection{{Real space approach from defect classification}}
\begin{figure}[bt]
\centering
\includegraphics[scale=0.65]{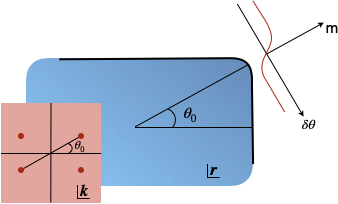}
\\
\caption{
An illustration of the defect Hamiltonian in Eq.~\eqref{eq:kxy}. We first restrict to the line joining two Dirac points related by $\ms{C}_2$ symmetry. Then within perturbation theory, it can be shown that the edge traps a point-like mass defect where the $\theta_0$-line intersects the edge of the sample in real-space.}
\label{Fig:without_C4}
\end{figure}

We consider a system with rounded corners, whose size is much greater than the lattice scale, as shown in Fig.~\ref{Fig:without_C4}, Without loss of generality, we take the BdG Hamiltonian
\begin{align}
H(\bm k)=& f_1 (\bm k) \sigma_x\tau_z + f_2(\bm k) \sigma_z\tau_z - \mu\tau_z\nonumber\\
&+ \epsilon g_1(\bm k) \tau_x + \epsilon g_2(\bm k) \tau_y \label{eq:1}
\end{align}
By the twofold rotation symmetry $\widehat{\ms{C}}_2=\widehat{\ms{C}}_4^2=\tau_z$, $f_{1,2}(\bm k)$ and $g_{1,2}(\bm k)$ are even and odd functions of $\bm k$, respectively. For convenience we take the weak pairing limit where  $\epsilon$ is small --- increasing $\epsilon$ will not change our result.
In the absence of additional spatial symmetries, we define the corner direction to be pointing from $\bm k=0$ to a pair of Dirac points, say at $\pm k_F(\cos\theta_0,\sin\theta_0)$. (Without loss of generality we assume a rectangular lattice.) The rounded corner is parametrized by an angle $\theta\in (0,\pi/2)\ni \theta_0$. %Our strategy is to consider the low energy Hamiltonian $h(k_\|,\theta)$ of the quasi-1d system that derives from `flattening" the edge, as shown in Fig. .... 
We define the \emph{local momentum coordinates} $(k_\|(\theta), k_\perp(\theta))$ as locally parallel and perpendicular to the piece of edge parametrized by $\theta$. In turn, one can express $k_x$ and $k_y$ in terms of the local $k$ coordinates and $\theta$ as
\begin{align}
k_x(\theta)=&k_\perp \cos\theta - k_\| \sin\theta \nonumber\\
k_y(\theta)=&k_\perp \sin\theta + k_\| \cos\theta 
\label{eq:kxy}
\end{align}
Let us focus on the slice of the bulk attached to the piece of edge at $\theta_0$. Since our goal is to obtain the low-energy modes on a smooth edge, we take a continuum limit and expand around small $k_\|$. We expand the Hamiltonian using the local $k$ coordinates at $\theta=\theta_0$.
%\begin{align}
%&H(k_\perp, k_\|; \theta_0) \nonumber\\
%=& f_1(k_\perp,k_\|=0) \sigma_x\tau_z + \varphi_1(k_\perp)k_\|\sigma_x\tau_z \nonumber\\
%& + f_2(k_\perp,k_\|=0) \sigma_z\tau_z + \varphi_2(k_\perp)k_\|\sigma_z\tau_z  \nonumber\\
%&+ \epsilon g_1(k_\perp,k_\|=0) \tau_x  + \epsilon \gamma_1(k_\perp) k_\| \tau_x \nonumber\\
%&+ \epsilon g_2(k_\perp,k_\|=0) \tau_y  + \epsilon \gamma_2(k_\perp) k_\| \tau_y -\mu\tau_z
%\end{align}
%
%To see this, 
Before we do so,
it is convenient to perform a unitary transformation on the Hamiltonian $H\to \tilde H= UHU^\dagger$, $\sigma_{x,z}\to\tilde\sigma_{x,z}(k_\perp(\theta_0))= U(k_\perp(\theta_0))\sigma_{x,z}U^\dagger (k_\perp(\theta_0))$, $\tau_{x,y}\to\tilde\tau_{x,y}(k_\perp(\theta_0))= U(k_\perp(\theta_0))\tau_{x,y}U^\dagger (k_\perp(\theta_0))$, where 
\begin{align}
\!\!\!U(k_\perp(\theta_0)) =& \exp \(\frac{i\sigma_y}2 \tan^{-1}\frac{f_2(k_\perp,k_\|=0)}{f_1(k_\perp,k_\|=0)}\)\times\nonumber\\
&\exp \(-\frac{i\tau_z}2 \tan^{-1}\frac{g_2(k_\perp,k_\|=0)}{g_1(k_\perp,k_\|=0)}\)\bigg|_{\theta_0}.
\end{align}
This transformation is not onsite (since its Fourier transform is not a $\delta$-function), but since it is periodic and smooth in $k_\perp$, ($\tan^{-1}\[f_2(\bm k)/f_1(\bm k)\]$ is smooth even at the Dirac point), it is \emph{exponentially} localized in real space. Thus, such a transformation does not change the exponential localization of wave packets.
After the transformation, the Hamiltonian takes a simple form at $k_\|=0$:
\be
\tilde H(k_\perp,k_\|=0;\theta_0)=\tilde f_1(k_\perp)  \sigma_x\tau_z  + \epsilon \tilde g_1(k_\perp) \tau_x  - \mu\tau_z 
\label{eq:1d}
\ee
where 
\begin{align}
\tilde f_1(k_\perp) = f_1(k_\perp,k_\|=0)\sqrt{1+\frac{f_2^2(k_\perp,k_\|=0)}{f_1^2(k_\perp,k_\|=0)} }\nonumber\\
\tilde g_1(k_\perp) = g_1(k_\perp,k_\|=0)\sqrt{1+\frac{g_2^2(k_\perp,k_\|=0)}{g_1^2(k_\perp,k_\|=0)} }
\end{align}
The Hamiltonian \eqref{eq:1d} describes two decoupled copies of 1d $p$-wave superconductor.  As long as $\mu$ does not exceed the band width, they are in the weak pairing phase with Fermi momenta at the two Dirac points. They give rise to two zero modes at the boundary, which, in a disk geometry, is the edge with well-defined $k_\|$. The edge zero mode wave function can be found via solving a differential equation with the replacement $k_\perp \to i\partial_{\perp}$ with the proper boundary conditions. This has been done in Ref.~\onlinecite{Roy_2004}. Following the results there, in the weak-pairing limit, the wave function come predominantly from the low energy components in the bulk, i.e., modes near the Fermi point. The $k$-space part of the wave function of the edge state is given by
\begin{align}
\psi(k_\perp) = &\mathcal{N}\int_0^{\infty} e^{ik_{\perp}x}  \sin(k_F x) e^{-\Delta x} \frac{dx}{2\pi} \nonumber\\
=& \frac{\mathcal{N}}{4\pi}\[\frac{1}{k_\perp+k_F+i\Delta}-\frac{1}{k_\perp-k_F+i\Delta}\]% \nonumber \\
%\approx& \frac{i}{4}\(\tilde\delta_{\Delta} (k_{\perp}+k_F)-\tilde\delta_{\Delta} (k_{\perp}-k_F)\),
\end{align}
where $\Delta=\epsilon\gamma_1(k_F)$ and $\mathcal{N}$ is a normalization factor. In the weak pairing limit, $|\psi(k_\perp)|^2$ indeed is strongly peaked at $|k-k_F|\lesssim \Delta$. The internal part of the wave function is an eigenstate of $\tilde\sigma_x\tilde\tau_y $. We can further expand the Hamiltonian for small $k_\|$ as
\begin{align}
\tilde H(k_\|, k_\perp;\theta_0)=& \tilde f_1(k_\perp)  \sigma_x\tau_z  + \epsilon \tilde g_1(k_\perp)  \tau_x  - \mu\tau_z 
 \nonumber\\
&+ \varphi_1(k_\perp) k_\|    \sigma_x\tau_z +  \varphi_2(k_\perp) k_\|    \sigma_z\tau_z \nonumber\\
& + \epsilon\gamma_1( k_\perp) k_\|   \tau_x+ \epsilon\gamma_2(k_\perp) k_\| \tau_y,
 \label{eq:1}
\end{align}
Importantly, since $f_{1,2}(\bm k)$ are even functions and $g_{1,2}(\bm k)$ are odd, the expansion coefficients $\varphi_{1,2}(k_\perp)$ are an odd functions and $\gamma_{1,2}(k_\|)$ are even. For small $k_\|$, one can solve the spectrum of $\tilde H$ by perturbation theory. For a small $k_\|$, the spectrum of \eqref{eq:1} can be found by perturbation theory
\begin{align}
h(k_\|, \theta_0) =& \int dk_\perp |\psi(k_\perp)|^2\hat P^{+}_\perp  H (k_\|, k_\perp;\theta_0) \hat P^{+}_\perp \nonumber\\
=&\epsilon \gamma_2(k_F)k_\| P_\perp^{+}\tilde\tau_y,
\label{eq:4}
\end{align}
where $P^{\pm}_\perp=(1\pm\tilde\sigma_x\tilde\tau_y)/2$ is the internal projection operator for the two edge states. We have used the fact that $P_\perp^{+}\tilde\tau_yP_\perp^{+} = P_\perp^{+}\tilde\tau_y$. Importantly we see that \emph{only} the $\gamma_2(k_\perp)$ term survives the projection: $\varphi_{1,2}$ terms drop out because they are odd in $k_\perp$, and $\gamma_1$ term gets projected out by $P^+_\perp$.

\medskip \noindent We also expand the edge hamiltonian as a function of coordinate $\delta\theta=\theta-\theta_0$, which we treat as a small quantity. For this edge, the parallel and perpendicular momenta related to those at $\theta_0$ via 
\begin{align}
k_{\|} (\theta) = k_{\|}(\theta_0)- k_\perp(\theta_0)\delta\theta\nonumber\\
k_{\perp} (\theta) = k_{\perp} (\theta_0)+k_\|(\theta_0) \delta\theta.
\end{align}
At $k_\|(\theta)=0$, we have, to leading order, $k_\|(\theta_0)=k_\perp(\theta_0)\delta\theta$, and $k_\perp(\theta_0) = k_\perp(\theta)$.
The Hamiltonian Eq.~\eqref{eq:1} can then be reexpressed with local $\bm k$-coordinates at $\theta$ as
\begin{align}
\tilde H(k_\|=0, k_\perp;\theta)=& \tilde f_1(k_\perp) \tilde \sigma_x\tau_z  + \epsilon \tilde g_1(k_\perp) \tilde \tau_x  - \mu\tau_z 
 \nonumber\\
&+ \varphi_1(k_\perp)  k_\perp\delta\theta  \tilde \sigma_x\tau_z +  \varphi_2(k_\perp)  k_\perp\delta\theta   \tilde \sigma_z\tau_z \nonumber\\
& + \epsilon\gamma_1( k_\perp) k_\perp\delta\theta  \tilde \tau_x+ \epsilon\gamma_2(k_\perp)  k_\perp\delta\theta \tilde\tau_y,
%H(k_\|=0, k_\perp;\theta)=&
% \tilde f_1(k_\perp) \tilde \sigma_x\tilde\tau_z + \tilde \varphi_2(k_\perp) \delta\theta k_\perp\tilde \sigma_z\tau_z  - \mu\tau_z\nonumber\\
%& + \epsilon \gamma_1(\bm k) k_{\perp}  \tilde \tau_x + \epsilon\gamma_2(\bm k) (\delta\theta k_\perp)\tilde\tau_y,
 \label{eq:1a}
\end{align}
Without the  terms $\propto\delta\theta$, this again describes two decoupled 1d topological superconductors, leading to edge zero modes.
The $\delta\theta$ terms can be treated perturbatively. We have
\begin{align}
h(k_\|=0, \theta_0+\delta\theta) =& \int dk_\perp |\psi(k_\perp)|^2\hat P^{+}_\perp  H (k_\|, k_\perp;\theta_0) \hat P^{+}_\perp \nonumber\\
=&  \tilde\varphi_2(k_F) k_F  \delta\theta  P_\perp^+\tilde\sigma_z\tau_z.
\label{eq:7}
\end{align}
 Now we see that only the $\varphi_2(k_\perp)k_\perp $ term survives the momentum and internal projection, since this term is now even in $k_\perp$. Combining Eqs.~(\ref{eq:4}, \ref{eq:7}), we get the edge hamiltonian near $\theta_0$ and with small $k_\|'$ 
\be
h(k_{\|}, \theta_0+\delta\theta) = \epsilon \gamma_2(k_F) k_\| (P_\perp^+\tilde\tau_y) +  \tilde\varphi_2(k_F) k_F \delta\theta (P_\perp^+\tilde\sigma_z\tau_z)
\ee
Since $P_\perp^+$ commutes with both $\tilde\tau_y$ and $\tilde\sigma_z\tau_z$, we have $\{P_\perp^+\tau_y,P_\perp^+\tilde\sigma_z\tau_z\}=0$. As $P_\perp^+$ projects the internal Hilbert space dimension from 4 to 2, we can represent the edge hamiltonian in the projected Hilbert space as
\be
h(k_{\|}, \theta_0+\delta\theta) = \alpha k_\| s_x + \beta \delta\theta s_y.
\label{eq:domainwall}
\ee
where $s_{x,y}$ are Pauli matrices. 

\medskip \noindent {By analogy with the 1d Jackiw-Rebbi model~\cite{jackiw-rebbi-1975}, the Hamiltonian in \eqref{eq:domainwall} hosts a Majorana zero mode localized at $\theta=\theta_0$. Indeed, for 1d system the $\mathbb{Z}_2$ defect classification is very simple, which is simply the relative sign of the mass term on the two sides of $\theta=\theta_0$, which is nontrivial in our case. This completes our proof; notice that even though the nontrivial topology comes from an edge Hamiltonian, it is ultimately determined by the bulk properties.}

\section{Summary}
\label{Sec:conclude}
\medskip \noindent {In this work, we  analyzed the second-order topology of a generic two-band doped Dirac semimetal in 2d with four Dirac nodes that are subject to ${p}+i{p}$ pairing.  We showed that this model realizes either a HOTSC$_2$ or a BOTSC$_2$ phase, depending on the presence of an additional $\ms{C}_4$ symmetry. The two topological superconducting phases are intimately connected, with the difference being whether the energy gap protecting the Majorana modes is from the bulk or the boundary. We showed that while the nested Wilson loop approach in general fails to capture the boundary topological obstruction, both intrinsic higher-order topology and boundary-obstructed topology are naturally captured in an alternative  defect classification approach. We thus establish the `Dirac+$({p}+i{p})$' as a low-energy criterion for TSC$_2$ phase, which can be viewed as an extension of the family tree of $p$-wave TSC's.}

\medskip \noindent {Based on our results, it would be interesting to search for chiral superconductivity in Dirac materials, such as the surface of topological crystalline insulators and graphene-based systems. Moreover, our result can be generalized to three dimensions to $p$-wave superconductivity {in} Weyl semimetals, which we leave to future work.}

\acknowledgements
We thank Tomas Bzdusek, Mao Lin, Taylor Hughes, and Titus Neupert for useful discussions. AT is funded by the European Union’s Horizon 2020 research and innovation program under the Marie Skłodowska-Curie grant agreement number 701647. YW and AJ are supported by startup funds at University of Florida.

\bibliography{HOTSc2}

\begin{thebibliography}{62}%
\makeatletter
\providecommand \@ifxundefined [1]{%
 \@ifx{#1\undefined}
}%
\providecommand \@ifnum [1]{%
 \ifnum #1\expandafter \@firstoftwo
 \else \expandafter \@secondoftwo
 \fi
}%
\providecommand \@ifx [1]{%
 \ifx #1\expandafter \@firstoftwo
 \else \expandafter \@secondoftwo
 \fi
}%
\providecommand \natexlab [1]{#1}%
\providecommand \enquote  [1]{``#1''}%
\providecommand \bibnamefont  [1]{#1}%
\providecommand \bibfnamefont [1]{#1}%
\providecommand \citenamefont [1]{#1}%
\providecommand \href@noop [0]{\@secondoftwo}%
\providecommand \href [0]{\begingroup \@sanitize@url \@href}%
\providecommand \@href[1]{\@@startlink{#1}\@@href}%
\providecommand \@@href[1]{\endgroup#1\@@endlink}%
\providecommand \@sanitize@url [0]{\catcode `\\12\catcode `\$12\catcode
  `\&12\catcode `\#12\catcode `\^12\catcode `\_12\catcode `\%12\relax}%
\providecommand \@@startlink[1]{}%
\providecommand \@@endlink[0]{}%
\providecommand \url  [0]{\begingroup\@sanitize@url \@url }%
\providecommand \@url [1]{\endgroup\@href {#1}{\urlprefix }}%
\providecommand \urlprefix  [0]{URL }%
\providecommand \Eprint [0]{\href }%
\providecommand \doibase [0]{http://dx.doi.org/}%
\providecommand \selectlanguage [0]{\@gobble}%
\providecommand \bibinfo  [0]{\@secondoftwo}%
\providecommand \bibfield  [0]{\@secondoftwo}%
\providecommand \translation [1]{[#1]}%
\providecommand \BibitemOpen [0]{}%
\providecommand \bibitemStop [0]{}%
\providecommand \bibitemNoStop [0]{.\EOS\space}%
\providecommand \EOS [0]{\spacefactor3000\relax}%
\providecommand \BibitemShut  [1]{\csname bibitem#1\endcsname}%
\let\auto@bib@innerbib\@empty
%</preamble>
\bibitem [{\citenamefont {Chiu}\ \emph {et~al.}(2016)\citenamefont {Chiu},
  \citenamefont {Teo}, \citenamefont {Schnyder},\ and\ \citenamefont
  {Ryu}}]{Ching-Kai_2016}%
  \BibitemOpen
  \bibfield  {author} {\bibinfo {author} {\bibfnamefont {C.-K.}\ \bibnamefont
  {Chiu}}, \bibinfo {author} {\bibfnamefont {J.~C.~Y.}\ \bibnamefont {Teo}},
  \bibinfo {author} {\bibfnamefont {A.~P.}\ \bibnamefont {Schnyder}}, \ and\
  \bibinfo {author} {\bibfnamefont {S.}~\bibnamefont {Ryu}},\ }\href {\doibase
  10.1103/RevModPhys.88.035005} {\bibfield  {journal} {\bibinfo  {journal}
  {Rev. Mod. Phys.}\ }\textbf {\bibinfo {volume} {88}},\ \bibinfo {pages}
  {035005} (\bibinfo {year} {2016})}\BibitemShut {NoStop}%
\bibitem [{\citenamefont {Ryu}\ \emph {et~al.}(2010)\citenamefont {Ryu},
  \citenamefont {Schnyder}, \citenamefont {Furusaki},\ and\ \citenamefont
  {Ludwig}}]{Ryu_2010}%
  \BibitemOpen
  \bibfield  {author} {\bibinfo {author} {\bibfnamefont {S.}~\bibnamefont
  {Ryu}}, \bibinfo {author} {\bibfnamefont {A.~P.}\ \bibnamefont {Schnyder}},
  \bibinfo {author} {\bibfnamefont {A.}~\bibnamefont {Furusaki}}, \ and\
  \bibinfo {author} {\bibfnamefont {A.~W.~W.}\ \bibnamefont {Ludwig}},\ }\href
  {\doibase 10.1088/1367-2630/12/6/065010} {\bibfield  {journal} {\bibinfo
  {journal} {New Journal of Physics}\ }\textbf {\bibinfo {volume} {12}},\
  \bibinfo {pages} {065010} (\bibinfo {year} {2010})}\BibitemShut {NoStop}%
\bibitem [{\citenamefont {Kitaev}(2009)}]{Kitaev_periodic_table}%
  \BibitemOpen
  \bibfield  {author} {\bibinfo {author} {\bibfnamefont {A.}~\bibnamefont
  {Kitaev}},\ }\href {\doibase 10.1063/1.3149495} {\bibfield  {journal}
  {\bibinfo  {journal} {AIP Conference Proceedings}\ }\textbf {\bibinfo
  {volume} {1134}},\ \bibinfo {pages} {22} (\bibinfo {year} {2009})},\ \Eprint
  {http://arxiv.org/abs/https://aip.scitation.org/doi/pdf/10.1063/1.3149495}
  {https://aip.scitation.org/doi/pdf/10.1063/1.3149495} \BibitemShut {NoStop}%
\bibitem [{\citenamefont {Chen}\ \emph {et~al.}(2013)\citenamefont {Chen},
  \citenamefont {Gu}, \citenamefont {Liu},\ and\ \citenamefont
  {Wen}}]{Chen_2013}%
  \BibitemOpen
  \bibfield  {author} {\bibinfo {author} {\bibfnamefont {X.}~\bibnamefont
  {Chen}}, \bibinfo {author} {\bibfnamefont {Z.-C.}\ \bibnamefont {Gu}},
  \bibinfo {author} {\bibfnamefont {Z.-X.}\ \bibnamefont {Liu}}, \ and\
  \bibinfo {author} {\bibfnamefont {X.-G.}\ \bibnamefont {Wen}},\ }\href
  {\doibase 10.1103/PhysRevB.87.155114} {\bibfield  {journal} {\bibinfo
  {journal} {Phys. Rev. B}\ }\textbf {\bibinfo {volume} {87}},\ \bibinfo
  {pages} {155114} (\bibinfo {year} {2013})}\BibitemShut {NoStop}%
\bibitem [{\citenamefont {Fu}\ \emph {et~al.}(2007)\citenamefont {Fu},
  \citenamefont {Kane},\ and\ \citenamefont {Mele}}]{Fu_Kane_Mele_2007}%
  \BibitemOpen
  \bibfield  {author} {\bibinfo {author} {\bibfnamefont {L.}~\bibnamefont
  {Fu}}, \bibinfo {author} {\bibfnamefont {C.~L.}\ \bibnamefont {Kane}}, \ and\
  \bibinfo {author} {\bibfnamefont {E.~J.}\ \bibnamefont {Mele}},\ }\href
  {\doibase 10.1103/PhysRevLett.98.106803} {\bibfield  {journal} {\bibinfo
  {journal} {Phys. Rev. Lett.}\ }\textbf {\bibinfo {volume} {98}},\ \bibinfo
  {pages} {106803} (\bibinfo {year} {2007})}\BibitemShut {NoStop}%
\bibitem [{\citenamefont {Kane}\ and\ \citenamefont
  {Mele}(2005{\natexlab{a}})}]{Kane_Mele_2005}%
  \BibitemOpen
  \bibfield  {author} {\bibinfo {author} {\bibfnamefont {C.~L.}\ \bibnamefont
  {Kane}}\ and\ \bibinfo {author} {\bibfnamefont {E.~J.}\ \bibnamefont
  {Mele}},\ }\href {\doibase 10.1103/PhysRevLett.95.226801} {\bibfield
  {journal} {\bibinfo  {journal} {Phys. Rev. Lett.}\ }\textbf {\bibinfo
  {volume} {95}},\ \bibinfo {pages} {226801} (\bibinfo {year}
  {2005}{\natexlab{a}})}\BibitemShut {NoStop}%
\bibitem [{\citenamefont {Kane}\ and\ \citenamefont
  {Mele}(2005{\natexlab{b}})}]{Kane_2005}%
  \BibitemOpen
  \bibfield  {author} {\bibinfo {author} {\bibfnamefont {C.~L.}\ \bibnamefont
  {Kane}}\ and\ \bibinfo {author} {\bibfnamefont {E.~J.}\ \bibnamefont
  {Mele}},\ }\href {\doibase 10.1103/PhysRevLett.95.146802} {\bibfield
  {journal} {\bibinfo  {journal} {Phys. Rev. Lett.}\ }\textbf {\bibinfo
  {volume} {95}},\ \bibinfo {pages} {146802} (\bibinfo {year}
  {2005}{\natexlab{b}})}\BibitemShut {NoStop}%
\bibitem [{\citenamefont {Kitaev}(2001)}]{Kitaev_2001}%
  \BibitemOpen
  \bibfield  {author} {\bibinfo {author} {\bibfnamefont {A.~Y.}\ \bibnamefont
  {Kitaev}},\ }\href {\doibase 10.1070/1063-7869/44/10s/s29} {\bibfield
  {journal} {\bibinfo  {journal} {Physics-Uspekhi}\ }\textbf {\bibinfo {volume}
  {44}},\ \bibinfo {pages} {131} (\bibinfo {year} {2001})}\BibitemShut
  {NoStop}%
\bibitem [{\citenamefont {Schindler}\ \emph {et~al.}(2018)\citenamefont
  {Schindler}, \citenamefont {Cook}, \citenamefont {Vergniory}, \citenamefont
  {Wang}, \citenamefont {Parkin}, \citenamefont {Bernevig},\ and\ \citenamefont
  {Neupert}}]{schindler2018higher}%
  \BibitemOpen
  \bibfield  {author} {\bibinfo {author} {\bibfnamefont {F.}~\bibnamefont
  {Schindler}}, \bibinfo {author} {\bibfnamefont {A.~M.}\ \bibnamefont {Cook}},
  \bibinfo {author} {\bibfnamefont {M.~G.}\ \bibnamefont {Vergniory}}, \bibinfo
  {author} {\bibfnamefont {Z.}~\bibnamefont {Wang}}, \bibinfo {author}
  {\bibfnamefont {S.~S.}\ \bibnamefont {Parkin}}, \bibinfo {author}
  {\bibfnamefont {B.~A.}\ \bibnamefont {Bernevig}}, \ and\ \bibinfo {author}
  {\bibfnamefont {T.}~\bibnamefont {Neupert}},\ }\href@noop {} {\bibfield
  {journal} {\bibinfo  {journal} {Science advances}\ }\textbf {\bibinfo
  {volume} {4}},\ \bibinfo {pages} {eaat0346} (\bibinfo {year}
  {2018})}\BibitemShut {NoStop}%
\bibitem [{\citenamefont {Benalcazar}\ \emph
  {et~al.}(2017{\natexlab{a}})\citenamefont {Benalcazar}, \citenamefont
  {Bernevig},\ and\ \citenamefont {Hughes}}]{benalcazar2017quantized}%
  \BibitemOpen
  \bibfield  {author} {\bibinfo {author} {\bibfnamefont {W.~A.}\ \bibnamefont
  {Benalcazar}}, \bibinfo {author} {\bibfnamefont {B.~A.}\ \bibnamefont
  {Bernevig}}, \ and\ \bibinfo {author} {\bibfnamefont {T.~L.}\ \bibnamefont
  {Hughes}},\ }\href@noop {} {\bibfield  {journal} {\bibinfo  {journal}
  {Science}\ }\textbf {\bibinfo {volume} {357}},\ \bibinfo {pages} {61}
  (\bibinfo {year} {2017}{\natexlab{a}})}\BibitemShut {NoStop}%
\bibitem [{\citenamefont {Benalcazar}\ \emph
  {et~al.}(2017{\natexlab{b}})\citenamefont {Benalcazar}, \citenamefont
  {Bernevig},\ and\ \citenamefont {Hughes}}]{benalcazar2017electric}%
  \BibitemOpen
  \bibfield  {author} {\bibinfo {author} {\bibfnamefont {W.~A.}\ \bibnamefont
  {Benalcazar}}, \bibinfo {author} {\bibfnamefont {B.~A.}\ \bibnamefont
  {Bernevig}}, \ and\ \bibinfo {author} {\bibfnamefont {T.~L.}\ \bibnamefont
  {Hughes}},\ }\href@noop {} {\bibfield  {journal} {\bibinfo  {journal}
  {Physical Review B}\ }\textbf {\bibinfo {volume} {96}},\ \bibinfo {pages}
  {245115} (\bibinfo {year} {2017}{\natexlab{b}})}\BibitemShut {NoStop}%
\bibitem [{\citenamefont {You}\ \emph {et~al.}(2018)\citenamefont {You},
  \citenamefont {Devakul}, \citenamefont {Burnell},\ and\ \citenamefont
  {Neupert}}]{you2018higher}%
  \BibitemOpen
  \bibfield  {author} {\bibinfo {author} {\bibfnamefont {Y.}~\bibnamefont
  {You}}, \bibinfo {author} {\bibfnamefont {T.}~\bibnamefont {Devakul}},
  \bibinfo {author} {\bibfnamefont {F.}~\bibnamefont {Burnell}}, \ and\
  \bibinfo {author} {\bibfnamefont {T.}~\bibnamefont {Neupert}},\ }\href@noop
  {} {\bibfield  {journal} {\bibinfo  {journal} {arXiv preprint
  arXiv:1807.09788}\ } (\bibinfo {year} {2018})}\BibitemShut {NoStop}%
\bibitem [{\citenamefont {Kunst}\ \emph {et~al.}(2018)\citenamefont {Kunst},
  \citenamefont {van Miert},\ and\ \citenamefont
  {Bergholtz}}]{kunst2018lattice}%
  \BibitemOpen
  \bibfield  {author} {\bibinfo {author} {\bibfnamefont {F.~K.}\ \bibnamefont
  {Kunst}}, \bibinfo {author} {\bibfnamefont {G.}~\bibnamefont {van Miert}}, \
  and\ \bibinfo {author} {\bibfnamefont {E.~J.}\ \bibnamefont {Bergholtz}},\
  }\href@noop {} {\bibfield  {journal} {\bibinfo  {journal} {Physical Review
  B}\ }\textbf {\bibinfo {volume} {97}},\ \bibinfo {pages} {241405} (\bibinfo
  {year} {2018})}\BibitemShut {NoStop}%
\bibitem [{\citenamefont {Song}\ \emph {et~al.}(2017)\citenamefont {Song},
  \citenamefont {Fang},\ and\ \citenamefont {Fang}}]{song2017d}%
  \BibitemOpen
  \bibfield  {author} {\bibinfo {author} {\bibfnamefont {Z.}~\bibnamefont
  {Song}}, \bibinfo {author} {\bibfnamefont {Z.}~\bibnamefont {Fang}}, \ and\
  \bibinfo {author} {\bibfnamefont {C.}~\bibnamefont {Fang}},\ }\href@noop {}
  {\bibfield  {journal} {\bibinfo  {journal} {Physical review letters}\
  }\textbf {\bibinfo {volume} {119}},\ \bibinfo {pages} {246402} (\bibinfo
  {year} {2017})}\BibitemShut {NoStop}%
\bibitem [{\citenamefont {Geier}\ \emph {et~al.}(2018)\citenamefont {Geier},
  \citenamefont {Trifunovic}, \citenamefont {Hoskam},\ and\ \citenamefont
  {Brouwer}}]{geier-2018}%
  \BibitemOpen
  \bibfield  {author} {\bibinfo {author} {\bibfnamefont {M.}~\bibnamefont
  {Geier}}, \bibinfo {author} {\bibfnamefont {L.}~\bibnamefont {Trifunovic}},
  \bibinfo {author} {\bibfnamefont {M.}~\bibnamefont {Hoskam}}, \ and\ \bibinfo
  {author} {\bibfnamefont {P.~W.}\ \bibnamefont {Brouwer}},\ }\href {\doibase
  10.1103/PhysRevB.97.205135} {\bibfield  {journal} {\bibinfo  {journal} {Phys.
  Rev. B}\ }\textbf {\bibinfo {volume} {97}},\ \bibinfo {pages} {205135}
  (\bibinfo {year} {2018})}\BibitemShut {NoStop}%
\bibitem [{\citenamefont {Wang}\ \emph
  {et~al.}(2018{\natexlab{a}})\citenamefont {Wang}, \citenamefont {Lin},\ and\
  \citenamefont {Hughes}}]{wang2018weak}%
  \BibitemOpen
  \bibfield  {author} {\bibinfo {author} {\bibfnamefont {Y.}~\bibnamefont
  {Wang}}, \bibinfo {author} {\bibfnamefont {M.}~\bibnamefont {Lin}}, \ and\
  \bibinfo {author} {\bibfnamefont {T.~L.}\ \bibnamefont {Hughes}},\
  }\href@noop {} {\bibfield  {journal} {\bibinfo  {journal} {Physical Review
  B}\ }\textbf {\bibinfo {volume} {98}},\ \bibinfo {pages} {165144} (\bibinfo
  {year} {2018}{\natexlab{a}})}\BibitemShut {NoStop}%
\bibitem [{\citenamefont {Ezawa}(2018)}]{ezawa2018minimal}%
  \BibitemOpen
  \bibfield  {author} {\bibinfo {author} {\bibfnamefont {M.}~\bibnamefont
  {Ezawa}},\ }\href@noop {} {\bibfield  {journal} {\bibinfo  {journal} {arXiv
  preprint arXiv:1801.00437}\ } (\bibinfo {year} {2018})}\BibitemShut {NoStop}%
\bibitem [{\citenamefont {Khalaf}(2018)}]{khalaf2018higher}%
  \BibitemOpen
  \bibfield  {author} {\bibinfo {author} {\bibfnamefont {E.}~\bibnamefont
  {Khalaf}},\ }\href@noop {} {\bibfield  {journal} {\bibinfo  {journal}
  {Physical Review B}\ }\textbf {\bibinfo {volume} {97}},\ \bibinfo {pages}
  {205136} (\bibinfo {year} {2018})}\BibitemShut {NoStop}%
\bibitem [{\citenamefont {Matsugatani}\ and\ \citenamefont
  {Watanabe}(2018)}]{matsugatani2018connecting}%
  \BibitemOpen
  \bibfield  {author} {\bibinfo {author} {\bibfnamefont {A.}~\bibnamefont
  {Matsugatani}}\ and\ \bibinfo {author} {\bibfnamefont {H.}~\bibnamefont
  {Watanabe}},\ }\href@noop {} {\bibfield  {journal} {\bibinfo  {journal}
  {Physical Review B}\ }\textbf {\bibinfo {volume} {98}},\ \bibinfo {pages}
  {205129} (\bibinfo {year} {2018})}\BibitemShut {NoStop}%
\bibitem [{\citenamefont {Lin}\ and\ \citenamefont
  {Hughes}(2018)}]{lin2018topological}%
  \BibitemOpen
  \bibfield  {author} {\bibinfo {author} {\bibfnamefont {M.}~\bibnamefont
  {Lin}}\ and\ \bibinfo {author} {\bibfnamefont {T.~L.}\ \bibnamefont
  {Hughes}},\ }\href@noop {} {\bibfield  {journal} {\bibinfo  {journal}
  {Physical Review B}\ }\textbf {\bibinfo {volume} {98}},\ \bibinfo {pages}
  {241103} (\bibinfo {year} {2018})}\BibitemShut {NoStop}%
\bibitem [{\citenamefont {Dwivedi}\ \emph {et~al.}(2018)\citenamefont
  {Dwivedi}, \citenamefont {Hickey}, \citenamefont {Eschmann},\ and\
  \citenamefont {Trebst}}]{dwivedi2018majorana}%
  \BibitemOpen
  \bibfield  {author} {\bibinfo {author} {\bibfnamefont {V.}~\bibnamefont
  {Dwivedi}}, \bibinfo {author} {\bibfnamefont {C.}~\bibnamefont {Hickey}},
  \bibinfo {author} {\bibfnamefont {T.}~\bibnamefont {Eschmann}}, \ and\
  \bibinfo {author} {\bibfnamefont {S.}~\bibnamefont {Trebst}},\ }\href@noop {}
  {\bibfield  {journal} {\bibinfo  {journal} {Physical Review B}\ }\textbf
  {\bibinfo {volume} {98}},\ \bibinfo {pages} {054432} (\bibinfo {year}
  {2018})}\BibitemShut {NoStop}%
\bibitem [{\citenamefont {Langbehn}\ \emph {et~al.}(2017)\citenamefont
  {Langbehn}, \citenamefont {Peng}, \citenamefont {Trifunovic}, \citenamefont
  {von Oppen},\ and\ \citenamefont {Brouwer}}]{langbehn2017reflection}%
  \BibitemOpen
  \bibfield  {author} {\bibinfo {author} {\bibfnamefont {J.}~\bibnamefont
  {Langbehn}}, \bibinfo {author} {\bibfnamefont {Y.}~\bibnamefont {Peng}},
  \bibinfo {author} {\bibfnamefont {L.}~\bibnamefont {Trifunovic}}, \bibinfo
  {author} {\bibfnamefont {F.}~\bibnamefont {von Oppen}}, \ and\ \bibinfo
  {author} {\bibfnamefont {P.~W.}\ \bibnamefont {Brouwer}},\ }\href@noop {}
  {\bibfield  {journal} {\bibinfo  {journal} {Physical review letters}\
  }\textbf {\bibinfo {volume} {119}},\ \bibinfo {pages} {246401} (\bibinfo
  {year} {2017})}\BibitemShut {NoStop}%
\bibitem [{\citenamefont {Parameswaran}\ and\ \citenamefont
  {Wan}(2017)}]{parameswaran2017topological}%
  \BibitemOpen
  \bibfield  {author} {\bibinfo {author} {\bibfnamefont {S.~A.}\ \bibnamefont
  {Parameswaran}}\ and\ \bibinfo {author} {\bibfnamefont {Y.}~\bibnamefont
  {Wan}},\ }\href@noop {} {\bibfield  {journal} {\bibinfo  {journal} {Physics}\
  }\textbf {\bibinfo {volume} {10}},\ \bibinfo {pages} {132} (\bibinfo {year}
  {2017})}\BibitemShut {NoStop}%
\bibitem [{\citenamefont {Wang}\ \emph
  {et~al.}(2018{\natexlab{b}})\citenamefont {Wang}, \citenamefont {Liu},
  \citenamefont {Lu},\ and\ \citenamefont {Zhang}}]{wang2018high}%
  \BibitemOpen
  \bibfield  {author} {\bibinfo {author} {\bibfnamefont {Q.}~\bibnamefont
  {Wang}}, \bibinfo {author} {\bibfnamefont {C.-C.}\ \bibnamefont {Liu}},
  \bibinfo {author} {\bibfnamefont {Y.-M.}\ \bibnamefont {Lu}}, \ and\ \bibinfo
  {author} {\bibfnamefont {F.}~\bibnamefont {Zhang}},\ }\href@noop {}
  {\bibfield  {journal} {\bibinfo  {journal} {Physical review letters}\
  }\textbf {\bibinfo {volume} {121}},\ \bibinfo {pages} {186801} (\bibinfo
  {year} {2018}{\natexlab{b}})}\BibitemShut {NoStop}%
\bibitem [{\citenamefont {Trifunovic}\ and\ \citenamefont
  {Brouwer}(2019)}]{trifunovic2019higher}%
  \BibitemOpen
  \bibfield  {author} {\bibinfo {author} {\bibfnamefont {L.}~\bibnamefont
  {Trifunovic}}\ and\ \bibinfo {author} {\bibfnamefont {P.~W.}\ \bibnamefont
  {Brouwer}},\ }\href@noop {} {\bibfield  {journal} {\bibinfo  {journal}
  {Physical Review X}\ }\textbf {\bibinfo {volume} {9}},\ \bibinfo {pages}
  {011012} (\bibinfo {year} {2019})}\BibitemShut {NoStop}%
\bibitem [{\citenamefont {{Tiwari}}\ \emph {et~al.}(2019)\citenamefont
  {{Tiwari}}, \citenamefont {{Li}}, \citenamefont {{Bernevig}}, \citenamefont
  {{Neupert}},\ and\ \citenamefont {{Parameswaran}}}]{Tiwari_2019}%
  \BibitemOpen
  \bibfield  {author} {\bibinfo {author} {\bibfnamefont {A.}~\bibnamefont
  {{Tiwari}}}, \bibinfo {author} {\bibfnamefont {M.-H.}\ \bibnamefont {{Li}}},
  \bibinfo {author} {\bibfnamefont {B.~A.}\ \bibnamefont {{Bernevig}}},
  \bibinfo {author} {\bibfnamefont {T.}~\bibnamefont {{Neupert}}}, \ and\
  \bibinfo {author} {\bibfnamefont {S.~A.}\ \bibnamefont {{Parameswaran}}},\
  }\href@noop {} {\bibfield  {journal} {\bibinfo  {journal} {arXiv e-prints}\
  ,\ \bibinfo {eid} {arXiv:1905.11421}} (\bibinfo {year} {2019})},\ \Eprint
  {http://arxiv.org/abs/1905.11421} {arXiv:1905.11421 [cond-mat.str-el]}
  \BibitemShut {NoStop}%
\bibitem [{\citenamefont {{Li}}\ \emph {et~al.}(2019)\citenamefont {{Li}},
  \citenamefont {{Zhu}}, \citenamefont {{Benalcazar}},\ and\ \citenamefont
  {{Hughes}}}]{Tianhe_2019}%
  \BibitemOpen
  \bibfield  {author} {\bibinfo {author} {\bibfnamefont {T.}~\bibnamefont
  {{Li}}}, \bibinfo {author} {\bibfnamefont {P.}~\bibnamefont {{Zhu}}},
  \bibinfo {author} {\bibfnamefont {W.~A.}\ \bibnamefont {{Benalcazar}}}, \
  and\ \bibinfo {author} {\bibfnamefont {T.~L.}\ \bibnamefont {{Hughes}}},\
  }\href@noop {} {\bibfield  {journal} {\bibinfo  {journal} {arXiv e-prints}\
  ,\ \bibinfo {eid} {arXiv:1906.02752}} (\bibinfo {year} {2019})},\ \Eprint
  {http://arxiv.org/abs/1906.02752} {arXiv:1906.02752 [cond-mat.mes-hall]}
  \BibitemShut {NoStop}%
\bibitem [{\citenamefont {{You}}(2019)}]{You_2019}%
  \BibitemOpen
  \bibfield  {author} {\bibinfo {author} {\bibfnamefont {Y.}~\bibnamefont
  {{You}}},\ }\href@noop {} {\bibfield  {journal} {\bibinfo  {journal} {arXiv
  e-prints}\ ,\ \bibinfo {eid} {arXiv:1908.04299}} (\bibinfo {year} {2019})},\
  \Eprint {http://arxiv.org/abs/1908.04299} {arXiv:1908.04299
  [cond-mat.str-el]} \BibitemShut {NoStop}%
\bibitem [{\citenamefont {C\ifmmode \u{a}\else \u{a}\fi{}lug\ifmmode~\u{a}\else
  \u{a}\fi{}ru}\ \emph {et~al.}(2019)\citenamefont {C\ifmmode \u{a}\else
  \u{a}\fi{}lug\ifmmode~\u{a}\else \u{a}\fi{}ru}, \citenamefont {Juri\ifmmode
  \check{c}\else \v{c}\fi{}i\ifmmode~\acute{c}\else \'{c}\fi{}},\ and\
  \citenamefont {Roy}}]{Roy_2019}%
  \BibitemOpen
  \bibfield  {author} {\bibinfo {author} {\bibfnamefont {D.}~\bibnamefont
  {C\ifmmode \u{a}\else \u{a}\fi{}lug\ifmmode~\u{a}\else \u{a}\fi{}ru}},
  \bibinfo {author} {\bibfnamefont {V.}~\bibnamefont {Juri\ifmmode
  \check{c}\else \v{c}\fi{}i\ifmmode~\acute{c}\else \'{c}\fi{}}}, \ and\
  \bibinfo {author} {\bibfnamefont {B.}~\bibnamefont {Roy}},\ }\href {\doibase
  10.1103/PhysRevB.99.041301} {\bibfield  {journal} {\bibinfo  {journal} {Phys.
  Rev. B}\ }\textbf {\bibinfo {volume} {99}},\ \bibinfo {pages} {041301}
  (\bibinfo {year} {2019})}\BibitemShut {NoStop}%
\bibitem [{\citenamefont {Ahn}\ and\ \citenamefont
  {Yang}(2020)}]{ahn-yang-2020}%
  \BibitemOpen
  \bibfield  {author} {\bibinfo {author} {\bibfnamefont {J.}~\bibnamefont
  {Ahn}}\ and\ \bibinfo {author} {\bibfnamefont {B.-J.}\ \bibnamefont {Yang}},\
  }\href {\doibase 10.1103/PhysRevResearch.2.012060} {\bibfield  {journal}
  {\bibinfo  {journal} {Phys. Rev. Research}\ }\textbf {\bibinfo {volume}
  {2}},\ \bibinfo {pages} {012060} (\bibinfo {year} {2020})}\BibitemShut
  {NoStop}%
\bibitem [{\citenamefont {{Roy}}(2020)}]{roy-2020}%
  \BibitemOpen
  \bibfield  {author} {\bibinfo {author} {\bibfnamefont {B.}~\bibnamefont
  {{Roy}}},\ }\href@noop {} {\bibfield  {journal} {\bibinfo  {journal} {arXiv
  e-prints}\ ,\ \bibinfo {eid} {arXiv:2003.12566}} (\bibinfo {year} {2020})},\
  \Eprint {http://arxiv.org/abs/2003.12566} {arXiv:2003.12566
  [cond-mat.mes-hall]} \BibitemShut {NoStop}%
\bibitem [{\citenamefont {Wu}\ \emph {et~al.}(2020)\citenamefont {Wu},
  \citenamefont {Benalcazar}, \citenamefont {Li}, \citenamefont {Thomale},
  \citenamefont {Liu},\ and\ \citenamefont {Hu}}]{hu-2020}%
  \BibitemOpen
  \bibfield  {author} {\bibinfo {author} {\bibfnamefont {X.}~\bibnamefont
  {Wu}}, \bibinfo {author} {\bibfnamefont {W.~A.}\ \bibnamefont {Benalcazar}},
  \bibinfo {author} {\bibfnamefont {Y.}~\bibnamefont {Li}}, \bibinfo {author}
  {\bibfnamefont {R.}~\bibnamefont {Thomale}}, \bibinfo {author} {\bibfnamefont
  {C.-X.}\ \bibnamefont {Liu}}, \ and\ \bibinfo {author} {\bibfnamefont
  {J.}~\bibnamefont {Hu}},\ }\href@noop {} {\enquote {\bibinfo {title}
  {Boundary-obstructed topological high-t$_c$ superconductivity in iron
  pnictides},}\ } (\bibinfo {year} {2020}),\ \Eprint
  {http://arxiv.org/abs/2003.12204} {arXiv:2003.12204 [cond-mat.supr-con]}
  \BibitemShut {NoStop}%
\bibitem [{\citenamefont {Zhang}\ \emph {et~al.}(2019)\citenamefont {Zhang},
  \citenamefont {Hsu},\ and\ \citenamefont {Sarma}}]{zhang2019higherorder}%
  \BibitemOpen
  \bibfield  {author} {\bibinfo {author} {\bibfnamefont {R.-X.}\ \bibnamefont
  {Zhang}}, \bibinfo {author} {\bibfnamefont {Y.-T.}\ \bibnamefont {Hsu}}, \
  and\ \bibinfo {author} {\bibfnamefont {S.~D.}\ \bibnamefont {Sarma}},\
  }\href@noop {} {\enquote {\bibinfo {title} {Higher-order topological dirac
  superconductors},}\ } (\bibinfo {year} {2019}),\ \Eprint
  {http://arxiv.org/abs/1909.07980} {arXiv:1909.07980 [cond-mat.mes-hall]}
  \BibitemShut {NoStop}%
\bibitem [{\citenamefont {Zhang}\ \emph {et~al.}(2020)\citenamefont {Zhang},
  \citenamefont {Sau},\ and\ \citenamefont {Sarma}}]{zhang2020kitaev}%
  \BibitemOpen
  \bibfield  {author} {\bibinfo {author} {\bibfnamefont {R.-X.}\ \bibnamefont
  {Zhang}}, \bibinfo {author} {\bibfnamefont {J.~D.}\ \bibnamefont {Sau}}, \
  and\ \bibinfo {author} {\bibfnamefont {S.~D.}\ \bibnamefont {Sarma}},\
  }\href@noop {} {\enquote {\bibinfo {title} {Kitaev building-block
  construction for higher-order topological superconductors},}\ } (\bibinfo
  {year} {2020}),\ \Eprint {http://arxiv.org/abs/2003.02559} {arXiv:2003.02559
  [cond-mat.supr-con]} \BibitemShut {NoStop}%
\bibitem [{\citenamefont {Vu}\ \emph {et~al.}(2020)\citenamefont {Vu},
  \citenamefont {Zhang},\ and\ \citenamefont
  {Sarma}}]{vu2020timereversalinvariant}%
  \BibitemOpen
  \bibfield  {author} {\bibinfo {author} {\bibfnamefont {D.}~\bibnamefont
  {Vu}}, \bibinfo {author} {\bibfnamefont {R.-X.}\ \bibnamefont {Zhang}}, \
  and\ \bibinfo {author} {\bibfnamefont {S.~D.}\ \bibnamefont {Sarma}},\
  }\href@noop {} {\enquote {\bibinfo {title} {Time-reversal-invariant
  $c_2$-symmetric higher-order topological superconductors},}\ } (\bibinfo
  {year} {2020}),\ \Eprint {http://arxiv.org/abs/2005.03679} {arXiv:2005.03679
  [cond-mat.supr-con]} \BibitemShut {NoStop}%
\bibitem [{\citenamefont {Khalaf}\ \emph {et~al.}(2019)\citenamefont {Khalaf},
  \citenamefont {Benalcazar}, \citenamefont {Hughes},\ and\ \citenamefont
  {Queiroz}}]{khalaf_2019}%
  \BibitemOpen
  \bibfield  {author} {\bibinfo {author} {\bibfnamefont {E.}~\bibnamefont
  {Khalaf}}, \bibinfo {author} {\bibfnamefont {W.~A.}\ \bibnamefont
  {Benalcazar}}, \bibinfo {author} {\bibfnamefont {T.~L.}\ \bibnamefont
  {Hughes}}, \ and\ \bibinfo {author} {\bibfnamefont {R.}~\bibnamefont
  {Queiroz}},\ }\href@noop {} {\enquote {\bibinfo {title} {Boundary-obstructed
  topological phases},}\ } (\bibinfo {year} {2019}),\ \Eprint
  {http://arxiv.org/abs/1908.00011} {arXiv:1908.00011 [cond-mat.mes-hall]}
  \BibitemShut {NoStop}%
\bibitem [{\citenamefont {Read}\ and\ \citenamefont
  {Green}(2000)}]{Read_Green}%
  \BibitemOpen
  \bibfield  {author} {\bibinfo {author} {\bibfnamefont {N.}~\bibnamefont
  {Read}}\ and\ \bibinfo {author} {\bibfnamefont {D.}~\bibnamefont {Green}},\
  }\href {\doibase 10.1103/PhysRevB.61.10267} {\bibfield  {journal} {\bibinfo
  {journal} {Phys. Rev. B}\ }\textbf {\bibinfo {volume} {61}},\ \bibinfo
  {pages} {10267} (\bibinfo {year} {2000})}\BibitemShut {NoStop}%
\bibitem [{\citenamefont {Ivanov}(2001)}]{PhysRevLett.86.268}%
  \BibitemOpen
  \bibfield  {author} {\bibinfo {author} {\bibfnamefont {D.~A.}\ \bibnamefont
  {Ivanov}},\ }\href {\doibase 10.1103/PhysRevLett.86.268} {\bibfield
  {journal} {\bibinfo  {journal} {Phys. Rev. Lett.}\ }\textbf {\bibinfo
  {volume} {86}},\ \bibinfo {pages} {268} (\bibinfo {year} {2001})}\BibitemShut
  {NoStop}%
\bibitem [{\citenamefont {Lee}(1997)}]{RevModPhys.69.645}%
  \BibitemOpen
  \bibfield  {author} {\bibinfo {author} {\bibfnamefont {D.~M.}\ \bibnamefont
  {Lee}},\ }\href {\doibase 10.1103/RevModPhys.69.645} {\bibfield  {journal}
  {\bibinfo  {journal} {Rev. Mod. Phys.}\ }\textbf {\bibinfo {volume} {69}},\
  \bibinfo {pages} {645} (\bibinfo {year} {1997})}\BibitemShut {NoStop}%
\bibitem [{\citenamefont {Beenakker}(2013)}]{Beenakker}%
  \BibitemOpen
  \bibfield  {author} {\bibinfo {author} {\bibfnamefont {C.}~\bibnamefont
  {Beenakker}},\ }\href {\doibase 10.1146/annurev-conmatphys-030212-184337}
  {\bibfield  {journal} {\bibinfo  {journal} {Annual Review of Condensed Matter
  Physics}\ }\textbf {\bibinfo {volume} {4}},\ \bibinfo {pages} {113} (\bibinfo
  {year} {2013})},\ \Eprint
  {http://arxiv.org/abs/https://doi.org/10.1146/annurev-conmatphys-030212-184337}
  {https://doi.org/10.1146/annurev-conmatphys-030212-184337} \BibitemShut
  {NoStop}%
\bibitem [{\citenamefont {Qi}\ and\ \citenamefont
  {Zhang}(2011)}]{RevModPhys.83.1057}%
  \BibitemOpen
  \bibfield  {author} {\bibinfo {author} {\bibfnamefont {X.-L.}\ \bibnamefont
  {Qi}}\ and\ \bibinfo {author} {\bibfnamefont {S.-C.}\ \bibnamefont {Zhang}},\
  }\href {\doibase 10.1103/RevModPhys.83.1057} {\bibfield  {journal} {\bibinfo
  {journal} {Rev. Mod. Phys.}\ }\textbf {\bibinfo {volume} {83}},\ \bibinfo
  {pages} {1057} (\bibinfo {year} {2011})}\BibitemShut {NoStop}%
\bibitem [{\citenamefont {Schnyder}\ \emph {et~al.}(2008)\citenamefont
  {Schnyder}, \citenamefont {Ryu}, \citenamefont {Furusaki},\ and\
  \citenamefont {Ludwig}}]{Schnyder_2008}%
  \BibitemOpen
  \bibfield  {author} {\bibinfo {author} {\bibfnamefont {A.~P.}\ \bibnamefont
  {Schnyder}}, \bibinfo {author} {\bibfnamefont {S.}~\bibnamefont {Ryu}},
  \bibinfo {author} {\bibfnamefont {A.}~\bibnamefont {Furusaki}}, \ and\
  \bibinfo {author} {\bibfnamefont {A.~W.~W.}\ \bibnamefont {Ludwig}},\ }\href
  {\doibase 10.1103/PhysRevB.78.195125} {\bibfield  {journal} {\bibinfo
  {journal} {Phys. Rev. B}\ }\textbf {\bibinfo {volume} {78}},\ \bibinfo
  {pages} {195125} (\bibinfo {year} {2008})}\BibitemShut {NoStop}%
\bibitem [{\citenamefont {Qi}\ \emph {et~al.}(2010)\citenamefont {Qi},
  \citenamefont {Hughes},\ and\ \citenamefont {Zhang}}]{Qi_Hughes_Zhang_2010}%
  \BibitemOpen
  \bibfield  {author} {\bibinfo {author} {\bibfnamefont {X.-L.}\ \bibnamefont
  {Qi}}, \bibinfo {author} {\bibfnamefont {T.~L.}\ \bibnamefont {Hughes}}, \
  and\ \bibinfo {author} {\bibfnamefont {S.-C.}\ \bibnamefont {Zhang}},\ }\href
  {\doibase 10.1103/PhysRevB.81.134508} {\bibfield  {journal} {\bibinfo
  {journal} {Phys. Rev. B}\ }\textbf {\bibinfo {volume} {81}},\ \bibinfo
  {pages} {134508} (\bibinfo {year} {2010})}\BibitemShut {NoStop}%
\bibitem [{\citenamefont {Wang}\ \emph
  {et~al.}(2018{\natexlab{c}})\citenamefont {Wang}, \citenamefont {Lin},\ and\
  \citenamefont {Hughes}}]{lin-wang-hughes-2018}%
  \BibitemOpen
  \bibfield  {author} {\bibinfo {author} {\bibfnamefont {Y.}~\bibnamefont
  {Wang}}, \bibinfo {author} {\bibfnamefont {M.}~\bibnamefont {Lin}}, \ and\
  \bibinfo {author} {\bibfnamefont {T.~L.}\ \bibnamefont {Hughes}},\ }\href
  {\doibase 10.1103/PhysRevB.98.165144} {\bibfield  {journal} {\bibinfo
  {journal} {Phys. Rev. B}\ }\textbf {\bibinfo {volume} {98}},\ \bibinfo
  {pages} {165144} (\bibinfo {year} {2018}{\natexlab{c}})}\BibitemShut
  {NoStop}%
\bibitem [{\citenamefont {Wu}\ \emph {et~al.}(2019)\citenamefont {Wu},
  \citenamefont {Yan},\ and\ \citenamefont {Huang}}]{huang-2019}%
  \BibitemOpen
  \bibfield  {author} {\bibinfo {author} {\bibfnamefont {Z.}~\bibnamefont
  {Wu}}, \bibinfo {author} {\bibfnamefont {Z.}~\bibnamefont {Yan}}, \ and\
  \bibinfo {author} {\bibfnamefont {W.}~\bibnamefont {Huang}},\ }\href
  {\doibase 10.1103/PhysRevB.99.020508} {\bibfield  {journal} {\bibinfo
  {journal} {Phys. Rev. B}\ }\textbf {\bibinfo {volume} {99}},\ \bibinfo
  {pages} {020508} (\bibinfo {year} {2019})}\BibitemShut {NoStop}%
\bibitem [{\citenamefont {Ono}\ \emph {et~al.}(2019)\citenamefont {Ono},
  \citenamefont {Yanase},\ and\ \citenamefont {Watanabe}}]{Ono_2019}%
  \BibitemOpen
  \bibfield  {author} {\bibinfo {author} {\bibfnamefont {S.}~\bibnamefont
  {Ono}}, \bibinfo {author} {\bibfnamefont {Y.}~\bibnamefont {Yanase}}, \ and\
  \bibinfo {author} {\bibfnamefont {H.}~\bibnamefont {Watanabe}},\ }\href
  {\doibase 10.1103/PhysRevResearch.1.013012} {\bibfield  {journal} {\bibinfo
  {journal} {Phys. Rev. Research}\ }\textbf {\bibinfo {volume} {1}},\ \bibinfo
  {pages} {013012} (\bibinfo {year} {2019})}\BibitemShut {NoStop}%
\bibitem [{\citenamefont {Skurativska}\ \emph
  {et~al.}(2020{\natexlab{a}})\citenamefont {Skurativska}, \citenamefont
  {Neupert},\ and\ \citenamefont {Fischer}}]{Nastya_2020}%
  \BibitemOpen
  \bibfield  {author} {\bibinfo {author} {\bibfnamefont {A.}~\bibnamefont
  {Skurativska}}, \bibinfo {author} {\bibfnamefont {T.}~\bibnamefont
  {Neupert}}, \ and\ \bibinfo {author} {\bibfnamefont {M.~H.}\ \bibnamefont
  {Fischer}},\ }\href {\doibase 10.1103/PhysRevResearch.2.013064} {\bibfield
  {journal} {\bibinfo  {journal} {Phys. Rev. Research}\ }\textbf {\bibinfo
  {volume} {2}},\ \bibinfo {pages} {013064} (\bibinfo {year}
  {2020}{\natexlab{a}})}\BibitemShut {NoStop}%
\bibitem [{\citenamefont {Geier}\ \emph {et~al.}(2019)\citenamefont {Geier},
  \citenamefont {Brouwer},\ and\ \citenamefont {Trifunovic}}]{geier2019}%
  \BibitemOpen
  \bibfield  {author} {\bibinfo {author} {\bibfnamefont {M.}~\bibnamefont
  {Geier}}, \bibinfo {author} {\bibfnamefont {P.~W.}\ \bibnamefont {Brouwer}},
  \ and\ \bibinfo {author} {\bibfnamefont {L.}~\bibnamefont {Trifunovic}},\
  }\href@noop {} {\enquote {\bibinfo {title} {Symmetry-based indicators for
  topological bogoliubov-de gennes hamiltonians},}\ } (\bibinfo {year}
  {2019}),\ \Eprint {http://arxiv.org/abs/1910.11271} {arXiv:1910.11271
  [cond-mat.mes-hall]} \BibitemShut {NoStop}%
\bibitem [{\citenamefont {Fu}(2011)}]{Fu_TCI_2011}%
  \BibitemOpen
  \bibfield  {author} {\bibinfo {author} {\bibfnamefont {L.}~\bibnamefont
  {Fu}},\ }\href {\doibase 10.1103/PhysRevLett.106.106802} {\bibfield
  {journal} {\bibinfo  {journal} {Phys. Rev. Lett.}\ }\textbf {\bibinfo
  {volume} {106}},\ \bibinfo {pages} {106802} (\bibinfo {year}
  {2011})}\BibitemShut {NoStop}%
\bibitem [{\citenamefont {{Khalaf}}\ \emph {et~al.}(2018)\citenamefont
  {{Khalaf}}, \citenamefont {{Po}}, \citenamefont {{Vishwanath}},\ and\
  \citenamefont {{Watanabe}}}]{Khalaf_2018}%
  \BibitemOpen
  \bibfield  {author} {\bibinfo {author} {\bibfnamefont {E.}~\bibnamefont
  {{Khalaf}}}, \bibinfo {author} {\bibfnamefont {H.~C.}\ \bibnamefont {{Po}}},
  \bibinfo {author} {\bibfnamefont {A.}~\bibnamefont {{Vishwanath}}}, \ and\
  \bibinfo {author} {\bibfnamefont {H.}~\bibnamefont {{Watanabe}}},\ }\href
  {\doibase 10.1103/PhysRevX.8.031070} {\bibfield  {journal} {\bibinfo
  {journal} {Physical Review X}\ }\textbf {\bibinfo {volume} {8}},\ \bibinfo
  {eid} {031070} (\bibinfo {year} {2018})},\ \Eprint
  {http://arxiv.org/abs/1711.11589} {arXiv:1711.11589 [cond-mat.str-el]}
  \BibitemShut {NoStop}%
\bibitem [{\citenamefont {Teo}\ and\ \citenamefont {Kane}(2010)}]{Teo_Kane}%
  \BibitemOpen
  \bibfield  {author} {\bibinfo {author} {\bibfnamefont {.~C.~Y.}\ \bibnamefont
  {Teo}}\ and\ \bibinfo {author} {\bibfnamefont {C.~L.}\ \bibnamefont {Kane}},\
  }\href {\doibase 10.1103/PhysRevB.82.115120} {\bibfield  {journal} {\bibinfo
  {journal} {Phys. Rev. B}\ }\textbf {\bibinfo {volume} {82}},\ \bibinfo
  {pages} {115120} (\bibinfo {year} {2010})}\BibitemShut {NoStop}%
\bibitem [{\citenamefont {Skurativska}\ \emph
  {et~al.}(2020{\natexlab{b}})\citenamefont {Skurativska}, \citenamefont
  {Neupert},\ and\ \citenamefont {Fischer}}]{neupert-fischer-2020}%
  \BibitemOpen
  \bibfield  {author} {\bibinfo {author} {\bibfnamefont {A.}~\bibnamefont
  {Skurativska}}, \bibinfo {author} {\bibfnamefont {T.}~\bibnamefont
  {Neupert}}, \ and\ \bibinfo {author} {\bibfnamefont {M.~H.}\ \bibnamefont
  {Fischer}},\ }\href {\doibase 10.1103/PhysRevResearch.2.013064} {\bibfield
  {journal} {\bibinfo  {journal} {Phys. Rev. Research}\ }\textbf {\bibinfo
  {volume} {2}},\ \bibinfo {pages} {013064} (\bibinfo {year}
  {2020}{\natexlab{b}})}\BibitemShut {NoStop}%
\bibitem [{\citenamefont {Schindler}\ \emph {et~al.}(2020)\citenamefont
  {Schindler}, \citenamefont {Bradlyn}, \citenamefont {Fischer},\ and\
  \citenamefont {Neupert}}]{schindler-2020}%
  \BibitemOpen
  \bibfield  {author} {\bibinfo {author} {\bibfnamefont {F.}~\bibnamefont
  {Schindler}}, \bibinfo {author} {\bibfnamefont {B.}~\bibnamefont {Bradlyn}},
  \bibinfo {author} {\bibfnamefont {M.~H.}\ \bibnamefont {Fischer}}, \ and\
  \bibinfo {author} {\bibfnamefont {T.}~\bibnamefont {Neupert}},\ }\href@noop
  {} {\enquote {\bibinfo {title} {Pairing obstructions in topological
  superconductors},}\ } (\bibinfo {year} {2020}),\ \Eprint
  {http://arxiv.org/abs/2001.02682} {arXiv:2001.02682 [cond-mat.supr-con]}
  \BibitemShut {NoStop}%
\bibitem [{\citenamefont {Fang}\ \emph {et~al.}(2012)\citenamefont {Fang},
  \citenamefont {Gilbert},\ and\ \citenamefont {Bernevig}}]{Fang_2012}%
  \BibitemOpen
  \bibfield  {author} {\bibinfo {author} {\bibfnamefont {C.}~\bibnamefont
  {Fang}}, \bibinfo {author} {\bibfnamefont {M.~J.}\ \bibnamefont {Gilbert}}, \
  and\ \bibinfo {author} {\bibfnamefont {B.~A.}\ \bibnamefont {Bernevig}},\
  }\href {\doibase 10.1103/PhysRevB.86.115112} {\bibfield  {journal} {\bibinfo
  {journal} {Phys. Rev. B}\ }\textbf {\bibinfo {volume} {86}},\ \bibinfo
  {pages} {115112} (\bibinfo {year} {2012})}\BibitemShut {NoStop}%
\bibitem [{\citenamefont {Benalcazar}\ \emph {et~al.}(2019)\citenamefont
  {Benalcazar}, \citenamefont {Li},\ and\ \citenamefont
  {Hughes}}]{Taylor_Filling}%
  \BibitemOpen
  \bibfield  {author} {\bibinfo {author} {\bibfnamefont {W.~A.}\ \bibnamefont
  {Benalcazar}}, \bibinfo {author} {\bibfnamefont {T.}~\bibnamefont {Li}}, \
  and\ \bibinfo {author} {\bibfnamefont {T.~L.}\ \bibnamefont {Hughes}},\
  }\href {\doibase 10.1103/PhysRevB.99.245151} {\bibfield  {journal} {\bibinfo
  {journal} {Phys. Rev. B}\ }\textbf {\bibinfo {volume} {99}},\ \bibinfo
  {pages} {245151} (\bibinfo {year} {2019})}\BibitemShut {NoStop}%
\bibitem [{\citenamefont {Fukui}\ and\ \citenamefont
  {Fujiwara}(2010)}]{Fukui_2010}%
  \BibitemOpen
  \bibfield  {author} {\bibinfo {author} {\bibfnamefont {T.}~\bibnamefont
  {Fukui}}\ and\ \bibinfo {author} {\bibfnamefont {T.}~\bibnamefont
  {Fujiwara}},\ }\href {\doibase 10.1103/PhysRevB.82.184536} {\bibfield
  {journal} {\bibinfo  {journal} {Phys. Rev. B}\ }\textbf {\bibinfo {volume}
  {82}},\ \bibinfo {pages} {184536} (\bibinfo {year} {2010})}\BibitemShut
  {NoStop}%
\bibitem [{\citenamefont {Shiozaki}\ \emph {et~al.}(2012)\citenamefont
  {Shiozaki}, \citenamefont {Fukui},\ and\ \citenamefont
  {Fujimoto}}]{Shiozaki_2012}%
  \BibitemOpen
  \bibfield  {author} {\bibinfo {author} {\bibfnamefont {K.}~\bibnamefont
  {Shiozaki}}, \bibinfo {author} {\bibfnamefont {T.}~\bibnamefont {Fukui}}, \
  and\ \bibinfo {author} {\bibfnamefont {S.}~\bibnamefont {Fujimoto}},\ }\href
  {\doibase 10.1103/PhysRevB.86.125405} {\bibfield  {journal} {\bibinfo
  {journal} {Phys. Rev. B}\ }\textbf {\bibinfo {volume} {86}},\ \bibinfo
  {pages} {125405} (\bibinfo {year} {2012})}\BibitemShut {NoStop}%
\bibitem [{\citenamefont {Fu}\ and\ \citenamefont {Kane}(2008)}]{fu-kane-2008}%
  \BibitemOpen
  \bibfield  {author} {\bibinfo {author} {\bibfnamefont {L.}~\bibnamefont
  {Fu}}\ and\ \bibinfo {author} {\bibfnamefont {C.~L.}\ \bibnamefont {Kane}},\
  }\href {\doibase 10.1103/PhysRevLett.100.096407} {\bibfield  {journal}
  {\bibinfo  {journal} {Phys. Rev. Lett.}\ }\textbf {\bibinfo {volume} {100}},\
  \bibinfo {pages} {096407} (\bibinfo {year} {2008})}\BibitemShut {NoStop}%
\bibitem [{Note1()}]{Note1}%
  \BibitemOpen
  \bibinfo {note} {One subtlety is that the eigenstates of the Wannier
  Hamiltonian is \protect \emph {not} a Wannier state, but rather a Bloch state
  with definite $k_x$ and $k_y$. Its eigenvalues, on the other hand, are
  independent on $k_x$ and correspond to the center position of the Wannier
  states. We have suppressed the $k_x$ argument in the definition of $\protect
  \mathaccentV {hat}05E\nu _x(k_y)$.}\BibitemShut {Stop}%
\bibitem [{\citenamefont {Qi}\ \emph {et~al.}(2008)\citenamefont {Qi},
  \citenamefont {Hughes},\ and\ \citenamefont {Zhang}}]{qi-hughes-zhang-2008}%
  \BibitemOpen
  \bibfield  {author} {\bibinfo {author} {\bibfnamefont {X.-L.}\ \bibnamefont
  {Qi}}, \bibinfo {author} {\bibfnamefont {T.~L.}\ \bibnamefont {Hughes}}, \
  and\ \bibinfo {author} {\bibfnamefont {S.-C.}\ \bibnamefont {Zhang}},\ }\href
  {\doibase 10.1103/PhysRevB.78.195424} {\bibfield  {journal} {\bibinfo
  {journal} {Phys. Rev. B}\ }\textbf {\bibinfo {volume} {78}},\ \bibinfo
  {pages} {195424} (\bibinfo {year} {2008})}\BibitemShut {NoStop}%
\bibitem [{\citenamefont {Stone}\ and\ \citenamefont {Roy}(2004)}]{Roy_2004}%
  \BibitemOpen
  \bibfield  {author} {\bibinfo {author} {\bibfnamefont {M.}~\bibnamefont
  {Stone}}\ and\ \bibinfo {author} {\bibfnamefont {R.}~\bibnamefont {Roy}},\
  }\href {\doibase 10.1103/PhysRevB.69.184511} {\bibfield  {journal} {\bibinfo
  {journal} {Phys. Rev. B}\ }\textbf {\bibinfo {volume} {69}},\ \bibinfo
  {pages} {184511} (\bibinfo {year} {2004})}\BibitemShut {NoStop}%
\bibitem [{\citenamefont {Jackiw}\ and\ \citenamefont
  {Rebbi}(1976)}]{jackiw-rebbi-1975}%
  \BibitemOpen
  \bibfield  {author} {\bibinfo {author} {\bibfnamefont {R.}~\bibnamefont
  {Jackiw}}\ and\ \bibinfo {author} {\bibfnamefont {C.}~\bibnamefont {Rebbi}},\
  }\href {\doibase 10.1103/PhysRevD.13.3398} {\bibfield  {journal} {\bibinfo
  {journal} {Phys. Rev. D}\ }\textbf {\bibinfo {volume} {13}},\ \bibinfo
  {pages} {3398} (\bibinfo {year} {1976})}\BibitemShut {NoStop}%
\end{thebibliography}%
\end{document}